\documentclass[aps,prb,reprint,longbibliography]{revtex4-1}

\usepackage{graphicx}
\usepackage{amsmath}

\begin{document}

\title{Interference between the modes of an all-dielectric meta-atom}

\author{David~A.~Powell}
\email{david.a.powell@anu.edu.au}
\affiliation{Nonlinear Physics Centre and Centre for Ultrahigh-bandwidth Devices for Optical Systems (CUDOS),\\Research School of Physics and Engineering, The Australian National University, Canberra, ACT 2601, Australia.}

\begin{abstract}
The modes of silicon meta-atoms are investigated, motivated by their use as building blocks of Huygens' metasurfaces. A model based on these modes is presented, giving a clear physical explanation of all features in the extinction spectrum. Counter-intuitively, this can show negative contributions to extinction, which are shown to arise from the interference between non-orthogonal modes. The direct and interference contributions to extinction are determined, showing that conservation of energy is preserved. The Huygens' condition of matched electric and magnetic dipole moments leads to strong forward scattering and suppressed back scattering. It is shown that higher order modes with appropriate symmetry generalise this condition, leading to multiple bands of directional scattering. The presented results are obtained using a robust approach to find the modes of nano-photonic scatterers, commonly referred to as quasi-normal modes. By utilising an integral formulation of Maxwell's equations, this work avoids the problem of normalising diverging far-fields, which other approaches require. The model and presented results are implemented in open-source code.
\end{abstract}

\maketitle

\section{Introduction}

Dielectric resonators have applications in microwave and optical frequency ranges, including antennas \cite{luk_dielectric_2003} and as building blocks of metamaterials \cite{zhao_experimental_2008,ahmadi_physical_2008,ginn_realizing_2012}, particularly impedance-matched Huygen's metasurfaces \cite{decker_high-efficiency_2015}. The results obtained in such structures are typically explained in terms of modes, determined from the fields at peaks or dips in the spectrum. However, these ad-hoc methods cannot resolve multiple modes which overlap spectrally, nor can they show how each mode contributes to the spectral response. To obtain a complete picture of the physics of such structures, it is necessary to find the modes independently, as eigensolutions satisfying Maxwell's equations with no incident field.

Approximate methods for finding the modes of dielectric resonators are known \cite{van_bladel_resonances_1975}, which usually assume that $\varepsilon \gg 1$. These methods are inaccurate for the moderate values of permittivity available at optical frequencies, and more sophisticated methods are needed to account for radiation effects. Open nanophotonic resonators such as meta-atoms, nano-antennas and oligomers are typically strongly radiative systems, where loss cannot be treated as a perturbation. In many nanophotonic systems, material dispersion and losses cannot be neglected, further complicating the problem of finding their modes.

In radiating and dissipative systems the modes have complex frequencies $s_n = j\omega_n + \Omega_n$, corresponding to damped oscillations of the form $\exp(\Omega_n t)\cos(\omega_n t)$, with $\Omega_n < 0$ [using the time convention $\exp(st)$ with $s=j\omega+\Omega$]. The corresponding modal fields $\mathbf{E}_n$ do not possess the orthogonality usually found in the modes of closed systems, and they are commonly referred to as quasi-normal modes \cite{ching_quasinormal-mode_1998}. They are particularly useful for solving dipole emission problems \cite{kristensen_modes_2014}, since they allow a mode volume to be defined for open cavities \cite{sauvan_theory_2013}. A significant practical difficulty is the requirement to normalize a mode with diverging far-fields \cite{kristensen_normalization_2015}.

A different perspective on the modes of scatterers can be found within the microwave engineering literature \cite{baum_emerging_1976}, originally motivated by time-domain radar problems. By using integral methods to solve Maxwell's equation, only currents on the scatterer need to be solved for, avoiding the need to explicitly handle the diverging far-fields. As it is based on finding the singularities of a scattering operator, this approach is referred to as the singularity expansion method (SEM). The field distributions corresponding to these singularities are identical to the quasi-normal modes at the complex frequencies of the singularities $j\omega_n + \Omega_n$. The key difference is that when solving scattering problems on the $j\omega$ axis, the fields in the SEM approach are reconstructed from the dyadic Green's function, which remains finite in the far-field. Thus the SEM avoids the most significant practical disadvantage of quasi-normal modes based on fields.

Recently it has been shown that the singularity expansion method can be applied to meta-atoms and plasmonic resonators \cite{zheng_line_2013,Powell2014,makitalo_modes_2014}, clearly identifying the modes which contribute to scattering and coupling problems. However, finding all modes within a region of the complex frequency plane requires an iterative procedure with multiple contour integrations\cite{zheng_implementation_2014}. This greatly increases the computational burden, and it remains unclear how robust this procedure is. In addition, it has not yet been demonstrated whether all spectral features can be explained by such a model, particularly the interference between non-orthogonal modes in the extinction spectrum and suppression of back-scattering corresponding to the Huygens condition\cite{pfeiffer_metamaterial_2013}.

In this work, a robust integral approach to finding modes of open resonators is demonstrated for several all-dielectric meta-atoms, based on the singularity expansion method.  In contrast to previous works, it is not limited to bodies of rotation \cite{glisson_evaluation_1983}. It is shown how this leads to a clear decomposition of the extinction spectrum of a silicon disk, automatically accounting for interference between the non-orthogonal modes. By performing a vector spherical harmonic decomposition of each mode, the unidirectional scattering behaviour is explained. It is shown that higher-order modes can also interfere to supress back-scattering, corresponding to the generalized Huygens' condition \cite{kruk_invited_2016}. Examples are also presented of structures with reduced symmetry, leading to bianisotropic and birefringent meta-atoms.

\section{Modelling approach} \label{sec:modelling}

In this work, quantities are described using the time convention $\exp(st)$, with $s=j\omega+\Omega$, so that the imaginary part of frequency gives the oscillation rate, and the real part gives the decay rate. A frequency-domain function $f(s)$ has a corresponding time-domain function $f(t)$ which can be obtained through the inverse Laplace transform $f(t)=\mathcal{L}^{-1}\left\{f(s)\right\}$. Physically observable quantities must be represented by a real function in the time domain, thus they must satisfy the constraint $f(s^*) = f^*(s)$ in the frequency domain.

\subsection{The modes of an open resonator} \label{sec:defining-modes}

An overview of the integral equation method used to solve Maxwell's equations is given in Appendix \ref{sec:integral_method}, based on the surface equivalence principle. This yields a frequency-dependent matrix $\mathbf{Z}(s)$, which describes the response of the scatterer to an arbitrary excitation field. The unknown current vector $\mathbf{I}$ excited by incident field vector $\mathbf{V}$ is
\begin{equation} \label{eq:impedance}
\mathbf{I}(s)=\mathbf{Z}^{-1}(s)\cdot \mathbf{V}(s).
\end{equation}
This equation could be solved numerically, as is done in many commercial software packages. More interestingly, it serves as the starting point for developing the model based on modes.

If the matrix $\mathbf{Z}^{-1}$ is singular at frequency $s_n = j\omega_n + \Omega_n$, then finite current $\mathbf{I}$ can be supported, without requiring any excitation source $\mathbf{V}$. This is similar to the well-known case of modes in a closed, lossless system, except that in an open system, mode frequencies must have some finite damping rate $\Omega_n$. The most important singularities are the pole frequencies $s_n$, where the impedance matrix satisfies the equations:
\begin{equation}
\mathbf{Z}(s_n)\cdot\mathbf{I}_n=0 \qquad\qquad \mathbf{K}_n\cdot\mathbf{Z}(s_n)=0
\end{equation}
for non-zero vectors $\mathbf{I}_n$ and $\mathbf{K}_n$. Physically, $\mathbf{I}_n$ corresponds to the current distribution of the mode, and $\mathbf{K}_n$ determines how well the mode is matched to the incident field. Due to geometric symmetry, many modes are degenerate, with several different eigenvectors $\mathbf{I}_n$ and $\mathbf{K}_n$ having the same pole location $s_n$. For example, the electric dipole mode of a sphere can be excited by $x$, $y$ or $z$ polarized fields, and this mode is triply degenerate.  

The poles of the impedance matrix are found by a contour integration procedure, with details given in Appendix \ref{sec:pole_search}.  Figure~\ref{fig:contour} illustrates such a contour, which is chosen to encompass all modes which are likely to be of interest. It is offset slightly from the $j\omega$ axis to eliminate any modes which do not couple to incident radiation, hence have $\Omega_n=0$. The desired radiating modes are shown by green crosses, and have $\Omega_n < 0$. Since currents must be real functions in the time domain, for each pole there is a corresponding complex conjugate pole at $-j\omega_n + \Omega_n$, shown in orange. As the poles and residues are just complex conjugates of those with positive $j\omega_n$, they can be found by symmetry, and do not need to be included within the contour. Note that some poles are over-damped, with $j\omega_n=0$, and these poles do not appear in conjugate pairs. The contour incorporates the $j\omega=0$ axis in order to capture these poles.

\begin{figure}
\includegraphics[width=0.9\columnwidth]{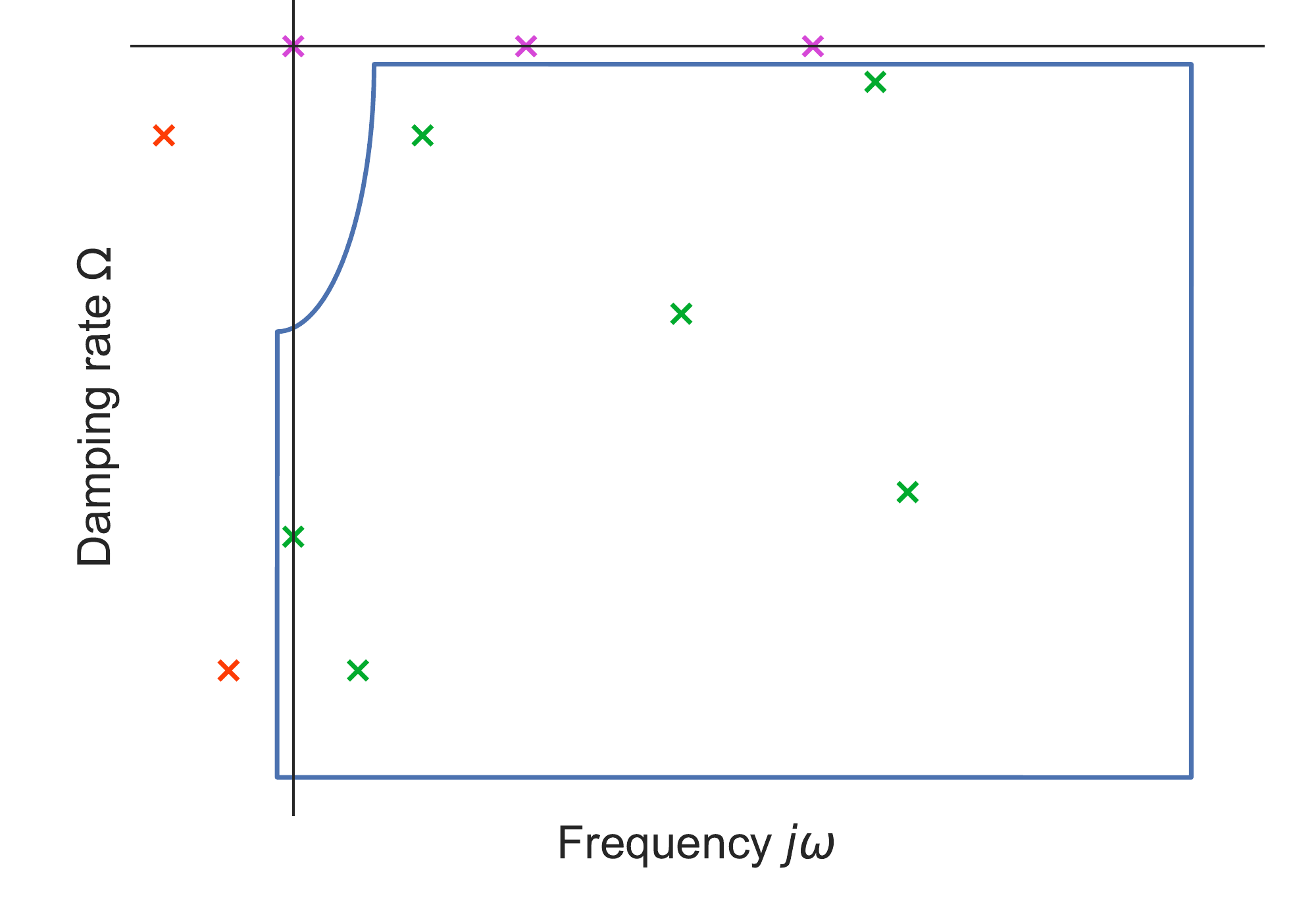}
\caption{Modes are found using a contour integration in the complex plane, which yields all enclosed poles $s_n = j\omega_n + \Omega_n$ and their residues with only a single integration. Green crosses: physical modes with finite radiation damping. Pink crosses: spurious internal solutions with no damping. Orange crosses: conjugate modes which can be found by symmetry.\label{fig:contour}}
\end{figure}

In general no orthogonality relation exists between the mode current vectors $\mathbf{I}_n$ and $\mathbf{K}_n$. As is discussed in Appendix \ref{sec:orthogonality}, orthogonality is not required for this approach. It will be shown in Section \ref{sec:disk} how this non-orthogonality leads to physically meaningful interference effects.

\subsection{Expanding currents in terms of modes}

Once the modes have been found, the current can be solved for arbitrary incident fields
\begin{equation} \label{eq:current_modes}
\mathbf{I}(j\omega) = 
\sum_n
\mathbf{I}_n
\left(\frac{1}{j\omega - s_n} + \frac{1}{s_n}\right)
\mathbf{K}_n \cdot \mathbf{V}(j\omega),
\end{equation}
where we consider excitation at physically realisable frequencies on the $j\omega$ axis. The vector $\mathbf{K}_n$ operates on the incident field $\mathbf{V}$ to give its overlap with the mode. The bracketed term accounts for how close the excitation frequency is to the mode's resonant frequency. Note that this polynomial has the correct asymptotic behaviour, thus improving the convergence and removing the need to include an entire function contribution\cite{pearson_evidence_1981}. The important result obtained from Eq.~\eqref{eq:current_modes} is a scalar weighting of each mode's current vector $\mathbf{I}_n$.

Regardless of whether it is calculated directly from Eq.~\eqref{eq:impedance} or as a superposition of modes from Eq.~\eqref{eq:current_modes}, the current vector $\mathbf{I}$ can give the surface current over the entire structure using Eq.~\eqref{eq:current_expansion}. This current distribution could then be used to calculate the total electric and magnetic fields. However, many quantities of physical interest such as scattering, radiation forces and torques can be calculated directly \cite{reid_efficient_2015} from the current vector $\mathbf{I}$. The quantity of most interest is the extinction cross-section
\begin{equation} \label{eq:extinction-direct}
\sigma_\mathrm{ext}=\mathrm{Re}\left[\mathbf{V}^{*}(j\omega)\cdot \mathbf{I}(j\omega)\right]\eta_0/|E_0|^2,
\end{equation}
giving the total work done by the incident fields on the currents in normalized form. Here $|E_0|$ is the electric field of the incident plane-wave. This quantity can be defined for each mode by substituting the mode's current and its weighting from Eq.~\eqref{eq:current_modes}, yielding
\begin{equation} \label{eq:extinction-modes}
\sigma_{\mathrm{ext},n}=\mathrm{Re}\left[\mathbf{V}^{*}(j\omega)\cdot \mathbf{I}_n\right]\eta_0/|E_0|^2,
\end{equation}

\section{Silicon disks}\label{sec:disk}

\begin{figure}[t]
\includegraphics[width=\columnwidth]{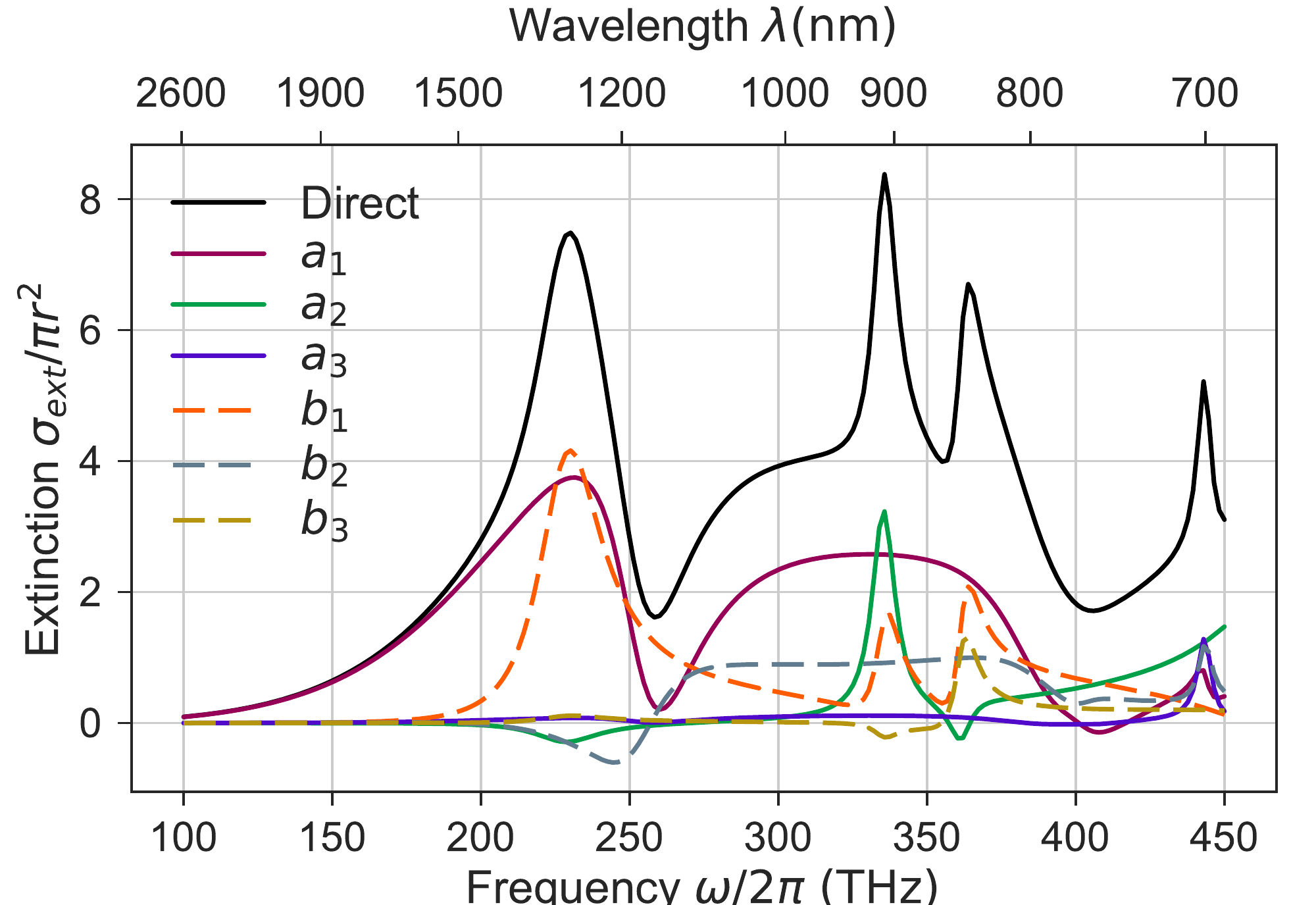}
\caption{The extinction cross-section of the the disk, direct calculation given by solid black line. Also shown is contributions from electric (solid) and magnetic (dashed) multipole moments. Curves are shown for different values of multipole order $l$, summed over all values of azimuthal index $m$.\label{fig:extinction_multipole}}
\end{figure}

The techniques outlined in Section \ref{sec:modelling} are now applied to study the scattering behavior of a single silicon disk meta-atom, an important building block of Huygens' metasurfaces. Initially, the structure is modelled directly using Eqs.~\eqref{eq:impedance} and \eqref{eq:extinction-direct}, without considering the modes.  The radius is taken as 242\,nm, height 220\,nm and edges are rounded with radius 50\,nm. The material properties of silicon were obtained by fitting an 8 pole model to the experimental data from Ref.~\onlinecite{green_optical_1995}. In Fig.~\ref{fig:extinction_multipole} the extinction cross-section of the disk is plotted by the solid black line. The incident wave-vector is parallel to the axis of the disk.

As a first attempt to explain the spectral features, a multipole expansion is also shown in Fig.~\ref{fig:extinction_multipole}. Details of the expansion are given in Appendix \ref{sec:multipole}. Solid lines show the electric multipole moments $a_l$, and dashed curves show the magnetic moments $b_l$. Although the multipoles accurately reproduce the total extinction, there is no direct correspondence between modes and multipoles, with each peak exhibiting contributions from many multipole moments. Furthermore, several multipole moments show peaks and dips at similar locations, but it is unclear if these moments are linked to each other. Therefore \emph{the multipole decomposition is unable to resolve the internal dynamics} which are observed in the extinction spectrum. It will be demonstrated that the model based on Eq.~\eqref{eq:current_modes} can resolve these internal dynamics, showing which modes correspond to each of the spectral features.

\subsection{Modes of the silicon disk}

\begin{figure}[tb]
\includegraphics[width=\columnwidth]{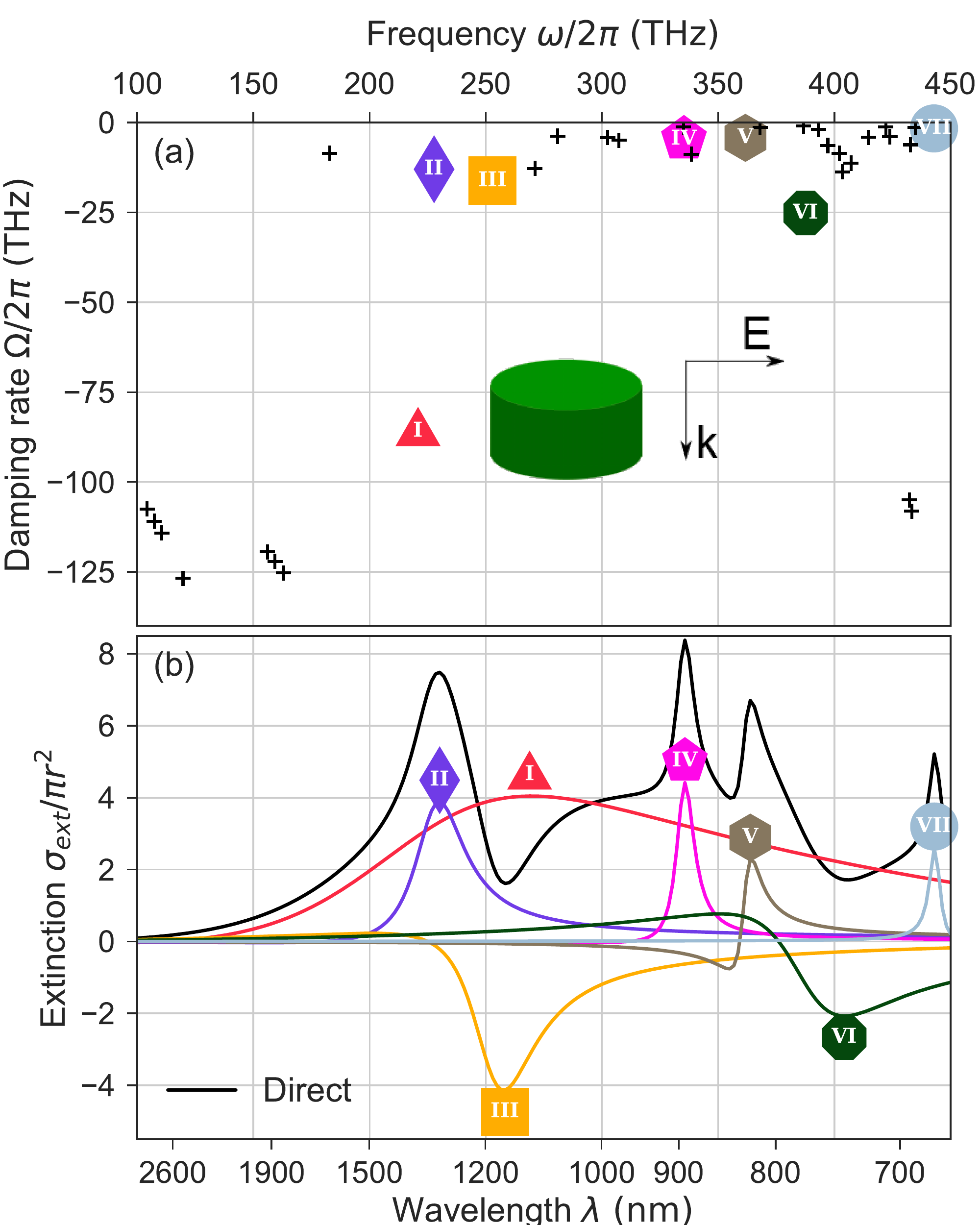}
\caption{(a) Complex frequencies of the modes of the silicon cylinder. Schematic shows the incident plane wave propagating along the cylinder axis. Coloured dots are the modes which couple strongly to the incident wave.
(b) Directly calculated extinction (black), and contributions from each of the modes. Colors indicate correspondence between poles and extinction curves.
\label{fig:extinction_modes}}
\end{figure}

\begin{figure}[tb!]
\centering
\includegraphics[width=0.95\columnwidth]{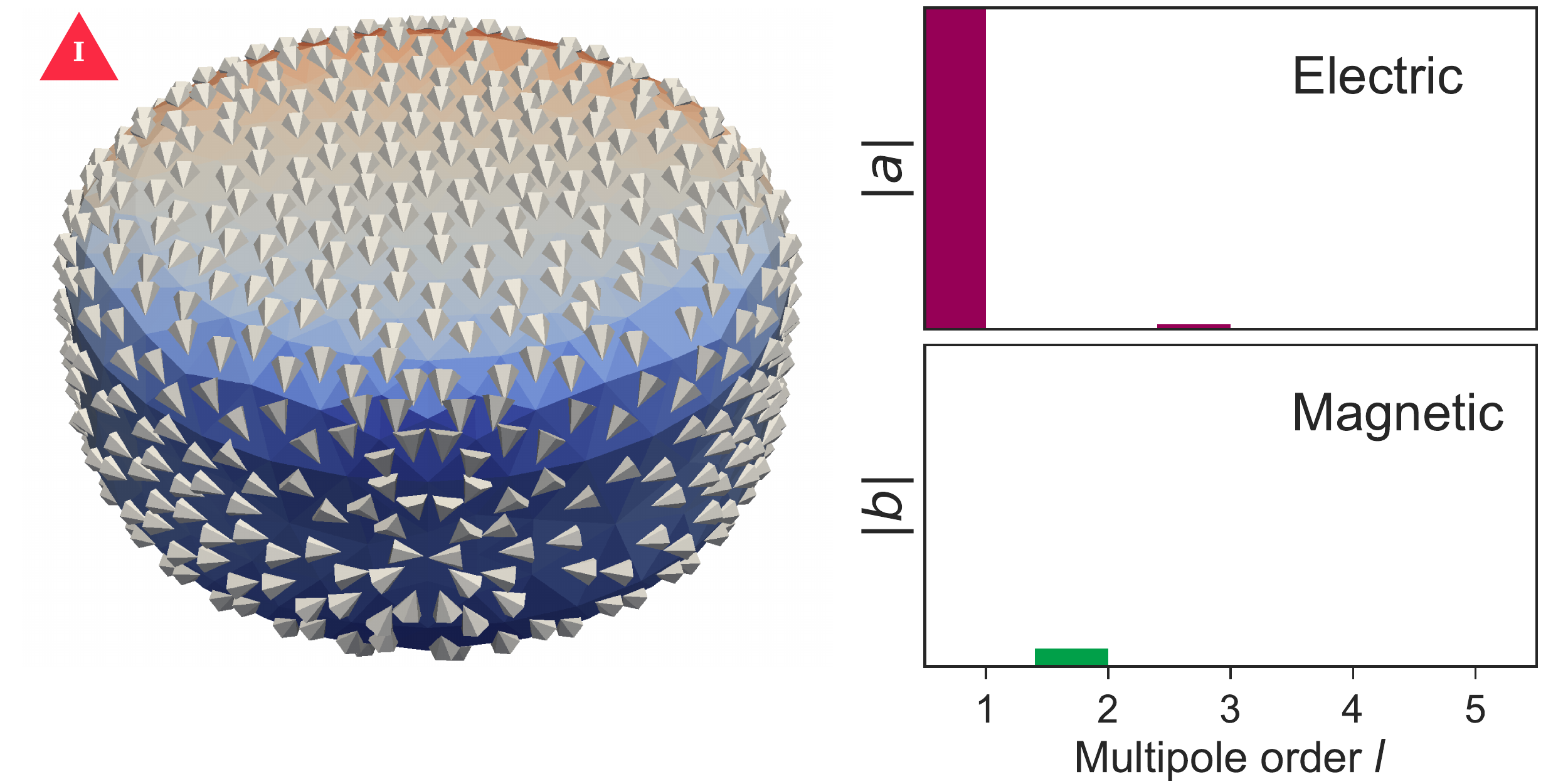}
\\[-10pt]\noindent\rule{0.95\columnwidth}{0.4pt}
\includegraphics[width=0.95\columnwidth]{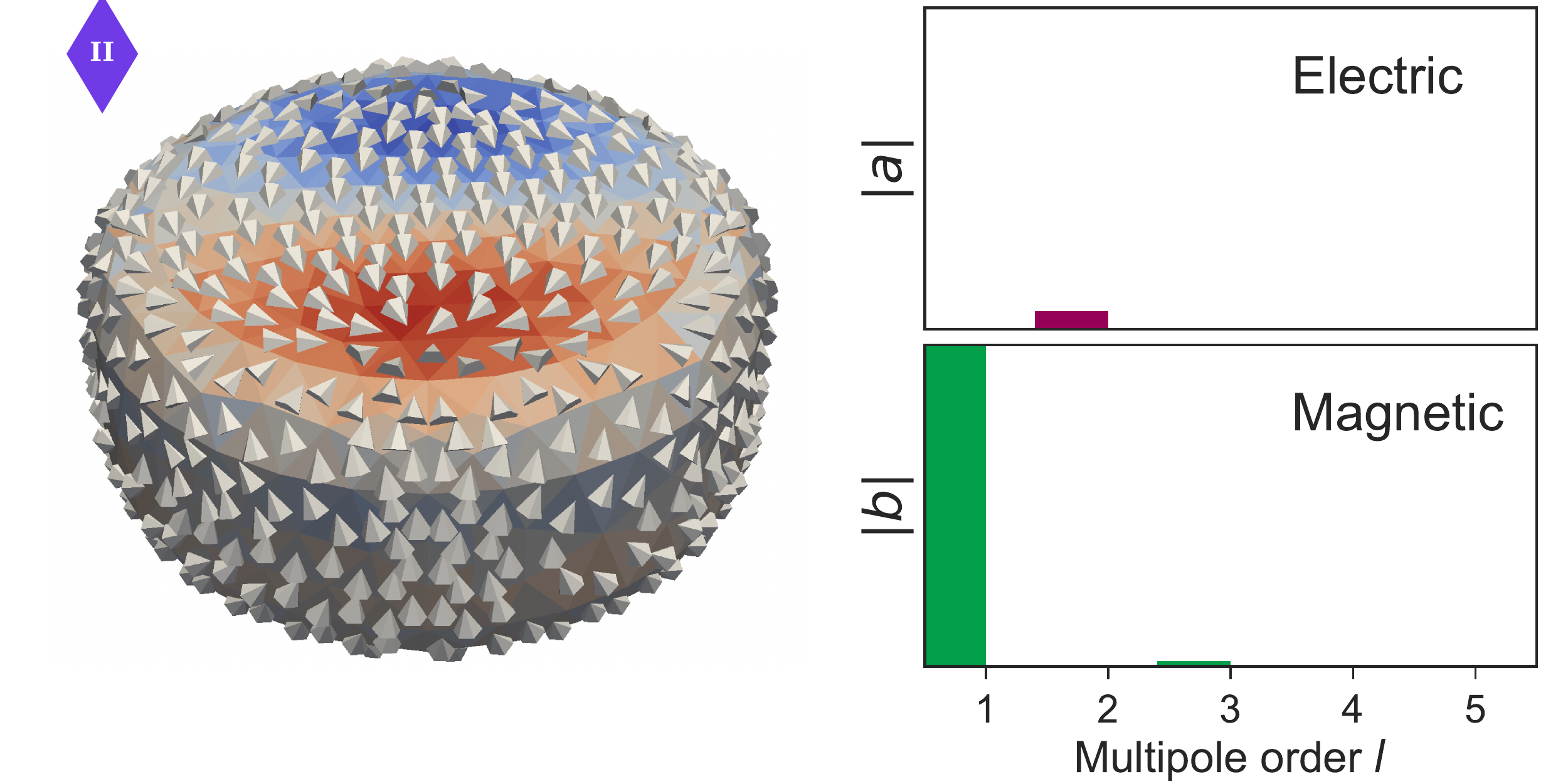}
\\[-10pt]\noindent\rule{0.95\columnwidth}{0.4pt}
\includegraphics[width=0.95\columnwidth]{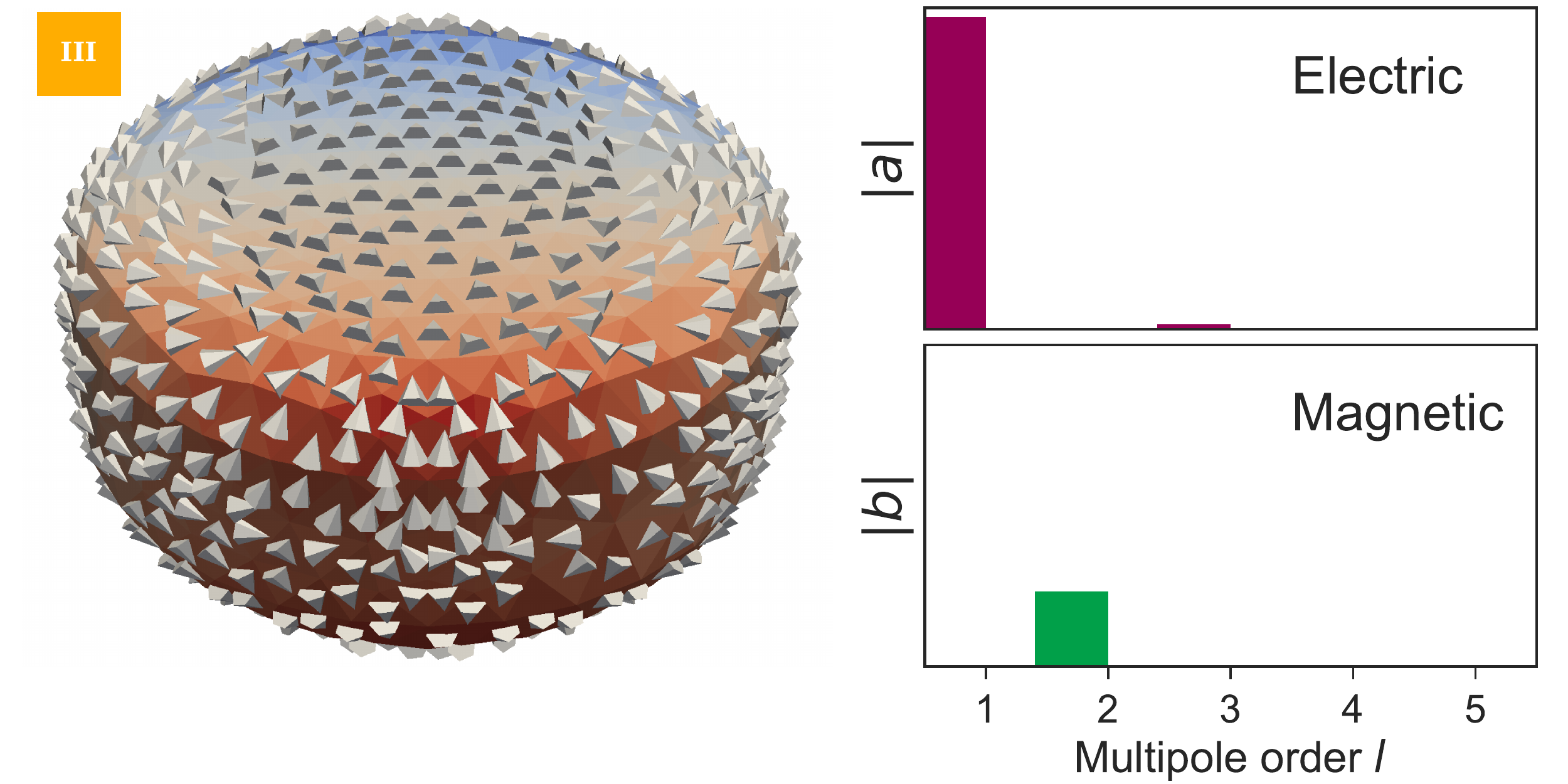}
\\[-10pt]\noindent\rule{0.95\columnwidth}{0.4pt}
\includegraphics[width=0.95\columnwidth]{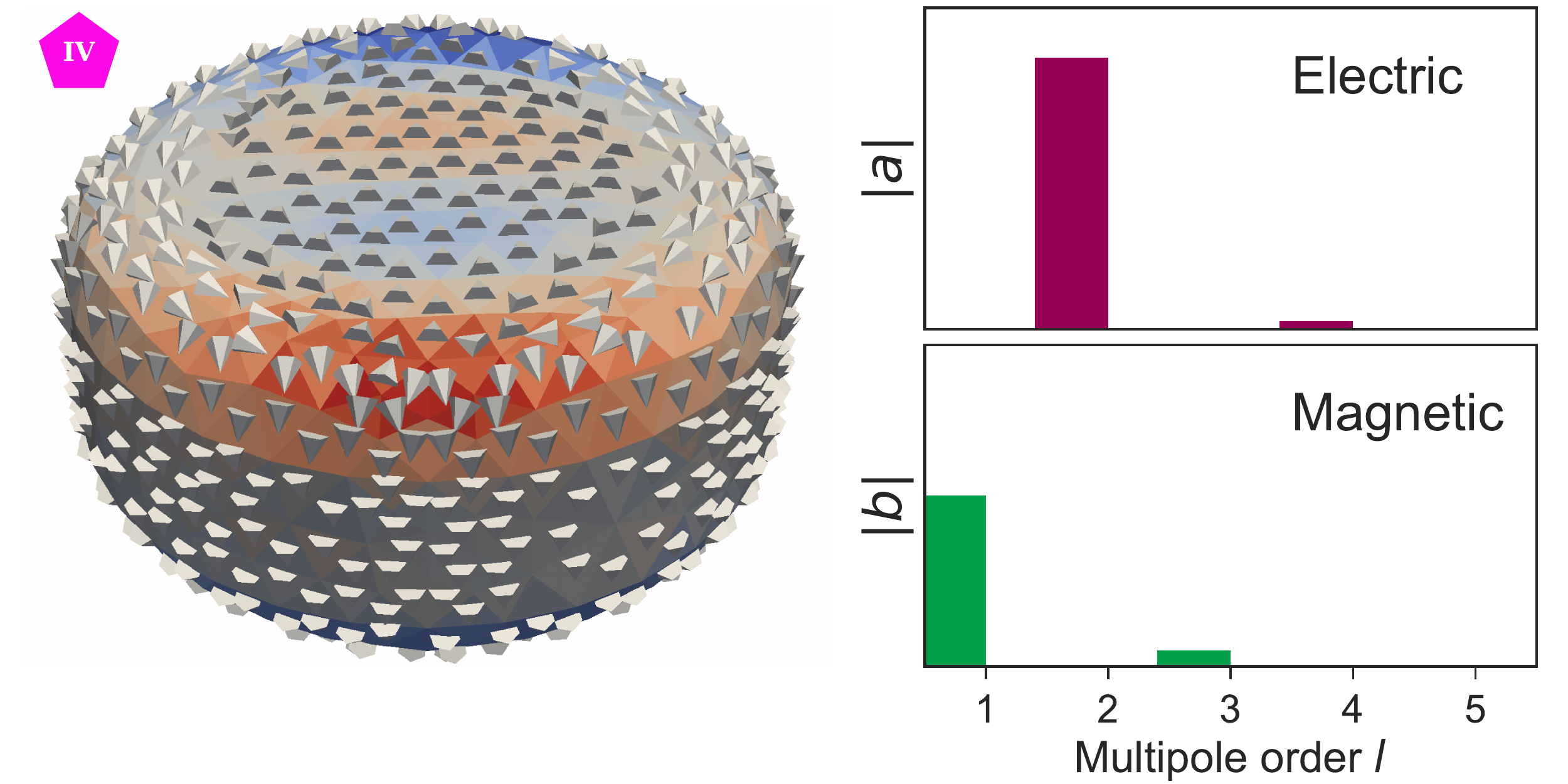}
\\[-10pt]\noindent\rule{0.95\columnwidth}{0.4pt}
\includegraphics[width=0.95\columnwidth]{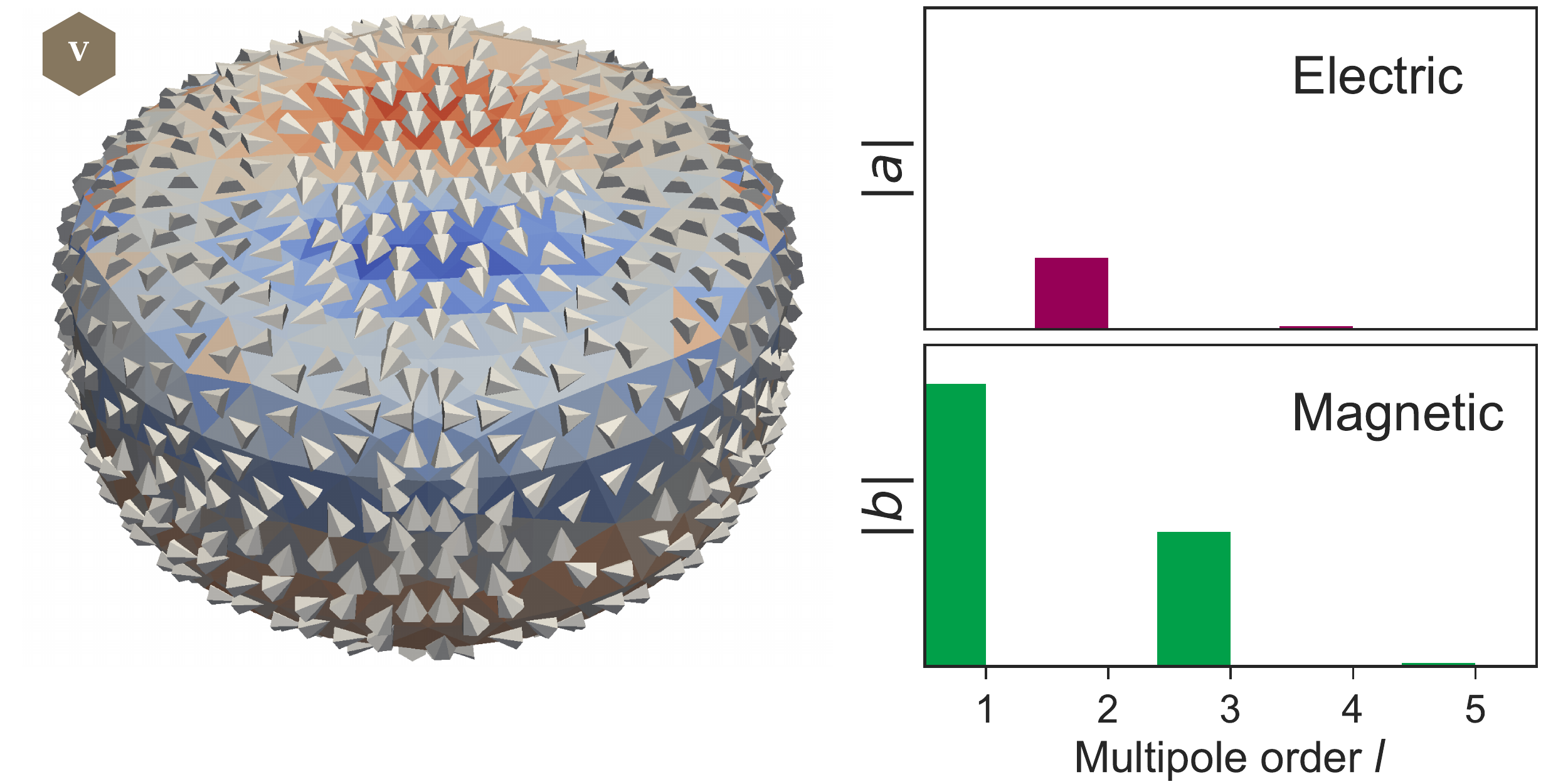}

\caption{Left: Modes of the disk, showing equivalent electric surface currents (arrows), and charges (colours). Markers correspond to poles in Fig.~\ref{fig:extinction_modes}. Right: Spherical multipoles of each mode, normalised to the total scattered power.\label{fig:mode-currents}}
\end{figure}

The modes of the silicon disk are found by the procedure outlined in Section \ref{sec:defining-modes} and Appendix \ref{sec:pole_search}. Figure \ref{fig:extinction_modes}(a) shows the location of the poles in the complex frequency plane, with many of them being doubly degenerate. Since currents decay in time as $e^{\Omega t}$, more highly damped modes have more negative values of $\Omega_n$. The schematic of the incident field orientation is shown in the inset. The modes which most strongly couple to this incident field are indicated with colored markers. The equivalent surface current $\mathbf{J}$ of the first 5 of these modes is shown in Fig.~\ref{fig:mode-currents}. Since these currents are complex, the plotted vectors give a snapshot of the oscillating current distribution. The divergence $\nabla\cdot\mathbf{J}$ is proportional to the equivalent surface charge (and hence to the normal component of the electric field) and is indicated by the shading of the surface. The colors of the markers next to each current distribution correspond to the poles shown in Fig.~\ref{fig:extinction_modes}(a). Each mode is also given an arbitrary label in Roman numerals for reference purposes. 

We can consider the dielectric disk to be a sphere which has been transformed in a continuous manner, breaking the spherical symmetry. By performing a multipole decomposition of the current for each mode of the disk, we can see which mode of the sphere it is most closely related to. This is shown in the right column of Fig.~\ref{fig:mode-currents}, where each mode's multipole moments are normalised to the total scattered power, as outlined in Appendix \ref{sec:multipole}. In all cases there is a single dominant multipole moment, although for higher order modes the influence of higher moments becomes more significant. In the following sections this multipole expansion of the modes will be used to explain their contributions to extinction and scattering.

Several of the modes shown in Fig.~\ref{fig:mode-currents} can be seen to correspond to well-known modes of cylindrical dielectric resonators shown in Refs.~\onlinecite{kajfez_computed_1984,mongia_dielectric_1994,ahmadi_physical_2008}. In particular, mode II is the $\mathrm{HEM}_{11\delta}$ mode (also known as $\mathrm{HE}_{11\delta}$) and mode III is the $\mathrm{HEM}_{12\delta}$ mode (also known as $\mathrm{EH}_{11\delta}$). The fundamental dipole-type mode I does not correspond to any of the modes presented in the cited works, but can be seen to closely resemble the $\mathrm{TE}_{111}$ mode of a closed metallic cavity \cite{kajfez_dielectric_1998}. It will be shown below that this mode makes a significant contribution to the response of the disk over a broad frequency range. 

\subsection{Extinction spectrum}

Figure~\ref{fig:extinction_modes}(b) shows the extinction contribution from each of the modes, calculated from Eqs.~\eqref{eq:current_modes} and \eqref{eq:extinction-modes}. The extinction from degenerate pairs of modes has been combined, along with the contribution of their conjugate modes at $-j\omega_n + \Omega_n$. All features in the extinction spectrum can be clearly attributed to the modal contributions. The extinction spectrum for each mode exhibits only a single feature, being a peak and/or dip in the vicinity of its pole frequency $\omega_n$. There is a very clear correspondence between the damping rate $\Omega_n$ and the sharpness of the features in the corresponding extinction curve. Note that for more highly damped modes, there is some shift between the peak and pole frequencies. This is because such modes couple strongly to the incident field, and therefore the overlap term in Eq.~\eqref{eq:current_modes} can shift the spectral features away from the natural frequency $j\omega_n$. The accuracy and convergence of this model of extinction is shown in Appendix \ref{sec:accuracy}.

One of the most striking features of Fig.~\ref{fig:extinction_modes}(b) is that several modes show negative contributions to extinction. This is due to the non-orthogonality of the modes, which means that even if the incident field matches the profile of one mode, it may still excite others. It can be seen that the dip in extinction at around 260\,THz can be attributed to a strong negative contribution from mode III, emitting radiation in the forward direction that is in-phase with the incident field. 

\begin{figure}[htb]
\includegraphics[width=\columnwidth]{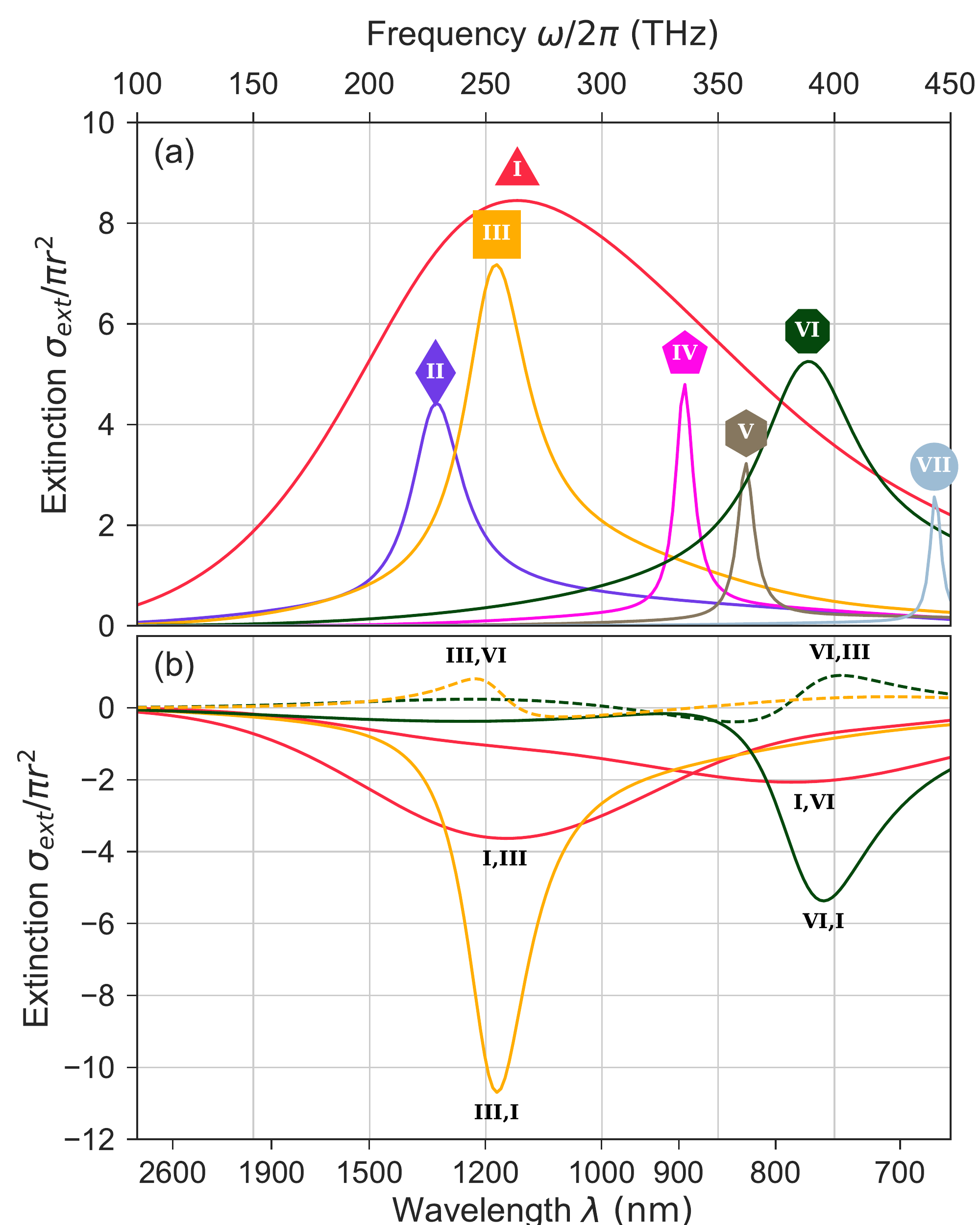}
\caption{Interference effects in the extinction spectrum of the silicon disk. (a) Direct extinction contribution $\sigma_{\mathrm{ext},n,n}$ of each mode. (b) The most significant interference terms between modes, $\sigma_{\mathrm{ext},m,n}$ for $m\neq n$ (solid lines), and selected weaker interference terms (dashed lines).
\label{fig:extinction-alternative}}
\end{figure}

To better illustrate this interference phenomenon, and to confirm that conservation of energy is not violated, the extinction is decomposed into direct terms from each mode, plus interference terms between every pair of modes\cite{hopkins_revisiting_2013}. This is done utilizing an alternative expression for extinction, based on the total rate of work done by the excited currents:
\begin{equation} \label{eq:extinction-alternate}
\sigma_\mathrm{ext}=\mathrm{Re}\left[\mathbf{I}^{*}(j\omega)\cdot \mathbf{Z}(j\omega)\cdot\mathbf{I}(j\omega)\right]\eta_0/|E_0|^2.
\end{equation}
This expression is quantifies the total power radiated and dissipated by the currents. Substituting Eq.~\eqref{eq:current_modes} into Eq.~\eqref{eq:extinction-alternate}, we can decompose the extinction into contributions from each pair of modes
\begin{equation} \label{eq:extinction-alternate-modes}
\sigma_{\mathrm{ext},m,n}=\mathrm{Re}\left[\mathbf{I}_m^{*}\cdot \mathbf{Z}(j\omega)\cdot\mathbf{I}_n\right]\eta_0/|E_0|^2.
\end{equation}
Here $\sigma_{\mathrm{ext},m,n}$ represents the rate of work done on the currents of mode $m$ by those of mode $n$. The self-terms $m=n$ represent the direct contribution of the mode to scattering and absorption, and must always be positive in a passive system. These terms are illustrated in Fig.~\ref{fig:extinction-alternative}(a), and it can be seen that they have much simpler line-shapes, and are positive as expected. Thus if any one of these modes was excited in isolation, there would be no negative contributions to extinction.

The off-diagonal terms $m\neq n$ explicitly show how modes $n$ and $m$ interact with each other. These are zero in a closed, lossless system with orthogonal modes, and in an open system they can also be zero for modes of opposite symmetry, such as modes I and II of the disk studied here. Fig.~\ref{fig:extinction-alternative}(b) shows the most significant interference terms for this structure. The conditions for significant interference between the modes are that they are non-orthogonal, and that they are both excited within the same spectral region. Thus we see that mode I, with its broad spectral response, interferes with both modes III and VI. On the other hand, although modes III and VI are also non-orthogonal, their limited spectral overlap gives much weaker interference, as is shown by the dashed curves. 

Due to passivity requirements, the interference terms between a pair of modes are constrained by the direct terms according to
\begin{equation} \label{eq:inteference-constraints}
\sigma_{\mathrm{ext},m,n} + \sigma_{\mathrm{ext},m,n} \geq -\left(\sigma_{\mathrm{ext},m,m} + \sigma_{\mathrm{ext},n,n}\right).
\end{equation}
The extinction obtained from Eq.~\eqref{eq:extinction-modes} can be understood as the sum of all direct and interference terms acting on mode $n$
\begin{equation}
\sigma_{\mathrm{ext},n} = \sum_m \sigma_{\mathrm{ext},n,m}.
\end{equation}
It should be emphasized that this summation does not need to be performed explicitly, and Eq.~\eqref{eq:current_modes} yields the total current coefficient for each mode accounting for all interference effects. This includes interference between any other modes which have not been explicitly incorporated within the model. Therefore a sufficient set of modes must be included within the model to have a physically meaningful result, otherwise Eq.~\eqref{eq:inteference-constraints} may be violated by some terms not being included.

\subsection{Total scattering}

\begin{figure}[b]
\includegraphics[width=\columnwidth]{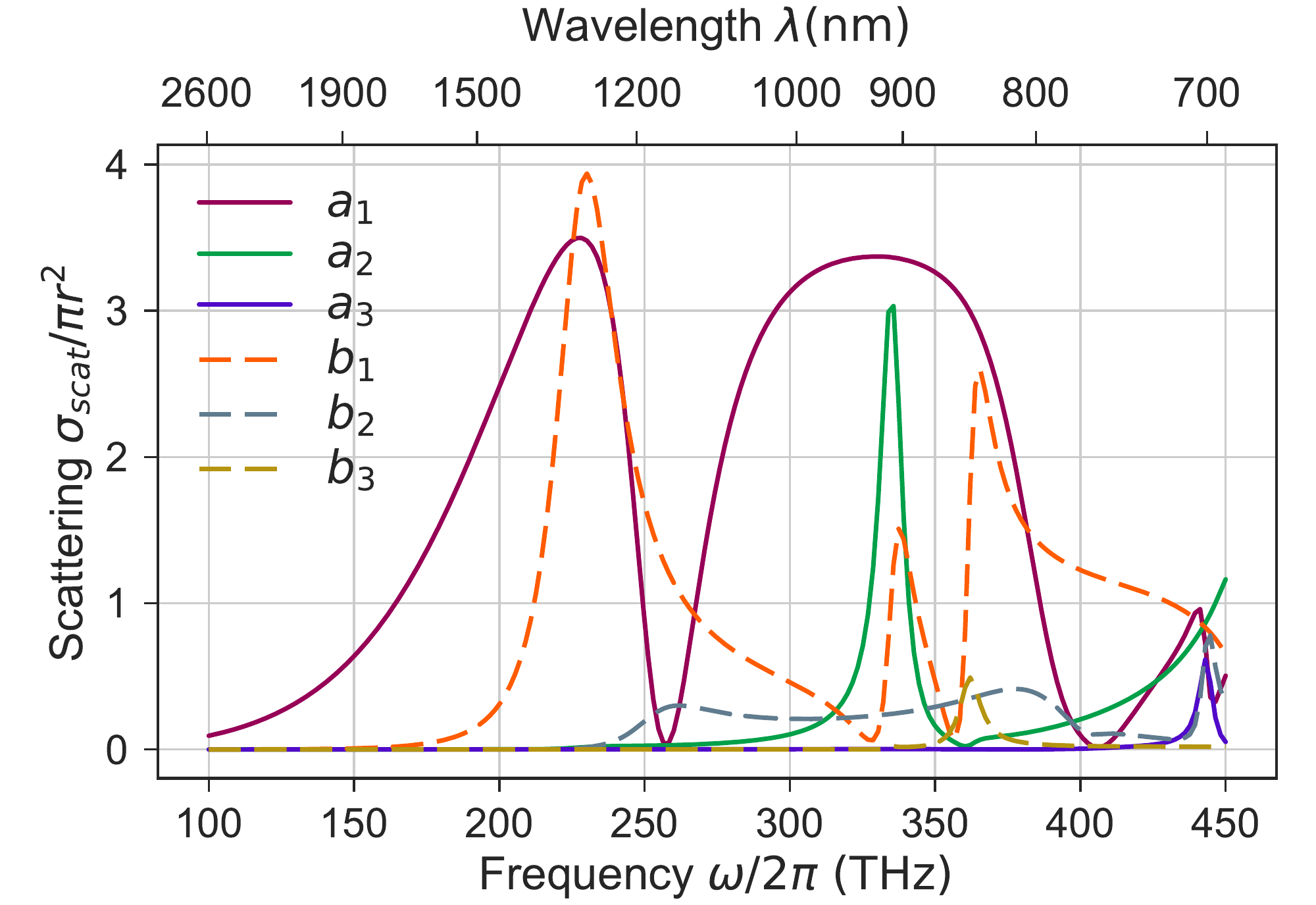}
\caption{Contribution of multipole moments to the scattering cross-section. \label{fig:multipole-scattering}}
\end{figure}

To calculate the total scattering cross section, vector spherical harmonics are used, since the total scattering is the incoherent sum of all multipole contributions, given by Eq.~\eqref{eq:multipole-scattering}. Figure \ref{fig:multipole-scattering} shows the contribution of each multipole coefficient to the scattering cross-section. As with the multipole extinction spectrum shown in Fig.~\ref{fig:extinction_multipole}, the features of the multipole scattering spectra are rather complex, but can be explained by considering the contributions of different modes. In the wavelength range above 1000\,nm, corresponding to the measured range in Ref.~\onlinecite{decker_high-efficiency_2015}, it can be seen that the scattering is dominated by the electric dipole and magnetic dipole  moments $a_{1}$ $b_{1}$. The magnetic dipole moment can be attributed to the resonance of mode II, which has negligible contributions from other moments.

The electric dipole moment $a_1$ appears to have two distinct maxima in Fig.~\ref{fig:multipole-scattering}. From the coefficients shown in Fig.~\ref{fig:mode-currents}, it is clear that only modes I and III contribute to this dipolar scattering. From Fig.~\ref{fig:extinction_modes}(b), we can see that mode I has a very broad resonance, while mode III has a much narrower resonance, with a negative contribution to extinction. This results in cancellation of electric dipole radiation, corresponding to an anapole distribution\cite{miroshnichenko_nonradiating_2015}. This effect is typically explained in terms of a quasi-static electric dipole (a linear current distribution) interfering with a toroidal dipole (a poloidal current distribution). 
The surface currents shown in Fig.~\ref{fig:mode-currents} are consistent with this picture, however the explanation in terms of modes is more general, and does not rely on any low frequency approximation. Indeed, in Ref.~\onlinecite{miroshnichenko_nonradiating_2015} it was shown that for spheres, the anapole distribution is excited when the contributions from the first and second $a_1$ modes cancel. The situation for the disk is similar, the difference being that the interfering modes I and III have additional contributions from other multipole moments.

\subsection{Directional scattering}

\begin{figure}[tb]
\includegraphics[width=\columnwidth]{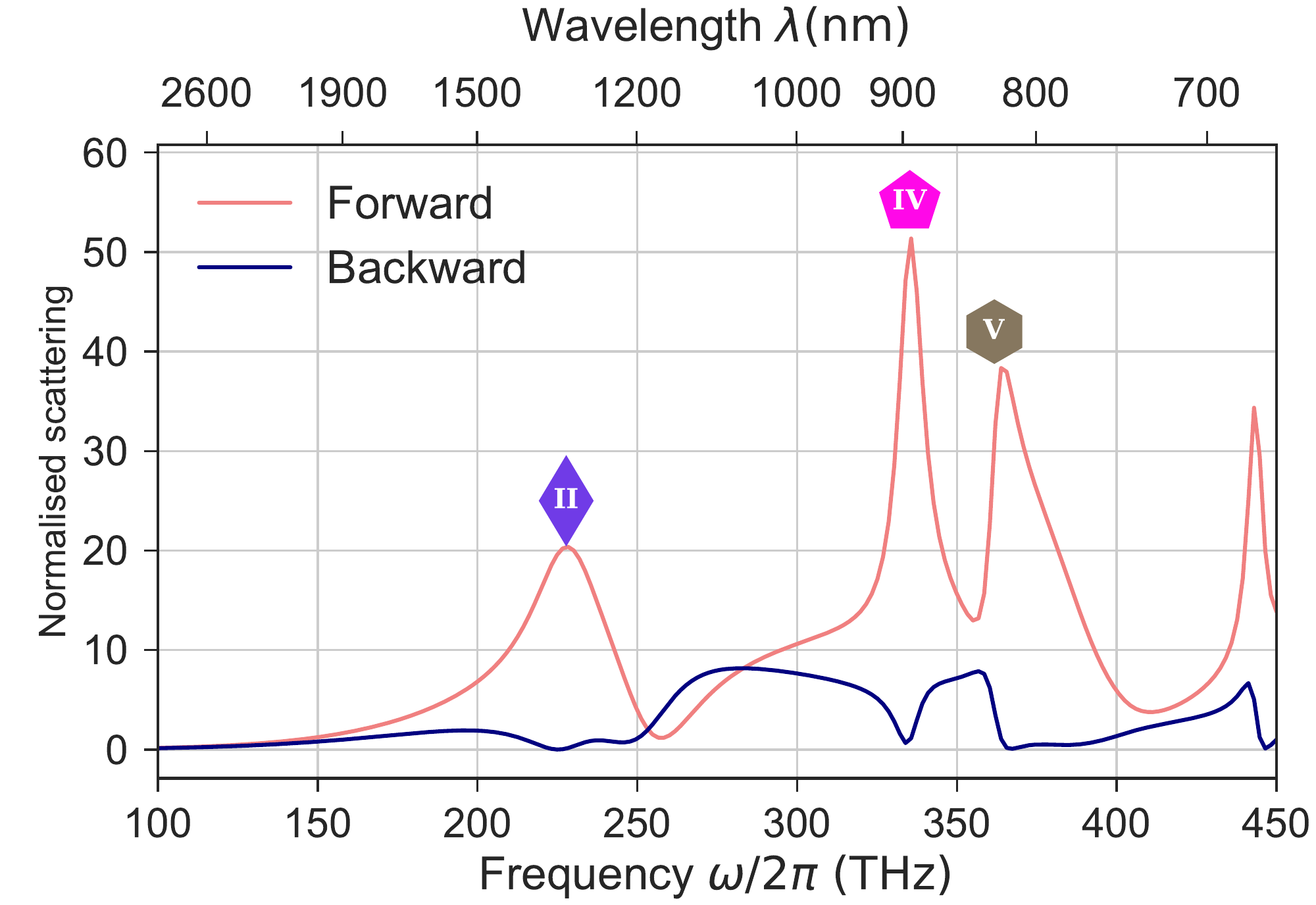}
\caption{Forward and backward scattering amplitudes. The markers indicate the modes corresponding to each of the peaks.\label{fig:directional-scattering}}
\end{figure}

For applications in Huygens' metasurfaces, the most important attribute of a meta-atom is to have suppressed back scattering and strong forward scattering. This is typically achieved by overlapping electric and magnetic dipole type resonances. Fig.~\ref{fig:directional-scattering} shows the forward and backward scattering amplitudes, with peaks labelled according to the corresponding resonant modes. The first peak of forward scattering corresponds to the overlap of modes I and III, with almost purely electric dipole radiation, and mode II, with almost purely magnetic dipole radiation. 

It can also be seen that at the resonances of modes IV and V there are additional highly directional scattering features, as these modes also overlap with the electric-dipole type modes I and III. Examining the multipole decompositions in Fig.~\ref{fig:mode-currents}, it can be seen that mode IV is dominated by its electric quadrupole response, with a significant contribution from its magnetic dipole response. In contrast, mode V is dominated by its magnetic dipole response, with lesser contributions from electric quadrupole and magnetic octupole moments. It is significant that all of these multipole moments radiate anti-symmetric electric fields into the forward and backward directions. When combined with the symmetric electric fields radiated by modes I and III, the backward scattering is cancelled, and the forward scattering is enhanced.

Considering the contribution of modes to this directional scattering process, the generalized Huygens condition introduced in Ref.~\onlinecite{kruk_invited_2016} can be re-interpreted as interference between modes of different symmetry. This suggests that to optimise this generalized Huygens' effect, the meta-atoms should be placed within a homogeneous dielectric environment \cite{decker_high-efficiency_2015}. A dielectric substrate without a compensating superstrate introduces bianisotropy by coupling modes of opposite symmetry\cite{Powell2010e}.

\section{Other structures}

The technique presented here is quite general, and can be applied to a variety of geometries. It is also applicable to a wide range of materials, as discussed in Appendix \ref{sec:materials}. The only significant limitation on geometry is that sharp corners need to be handled carefully, since they can cause numerical instability. The simplest solution is to round the edges with some finite radius, and such rounding is expected to occur in experimental samples. The approach makes no assumptions that the structure is smaller than the wavelength,  however as the structure becomes large compared to the wavelength, the number of modes typically increases quite dramatically, thus reducing the usefulness of the model. To demonstrate the generality of the method, it is applied to two additional structures.

\subsection{Bianisotropic disk}

\begin{figure}[tb!]
\centering
\includegraphics[width=0.94\columnwidth]{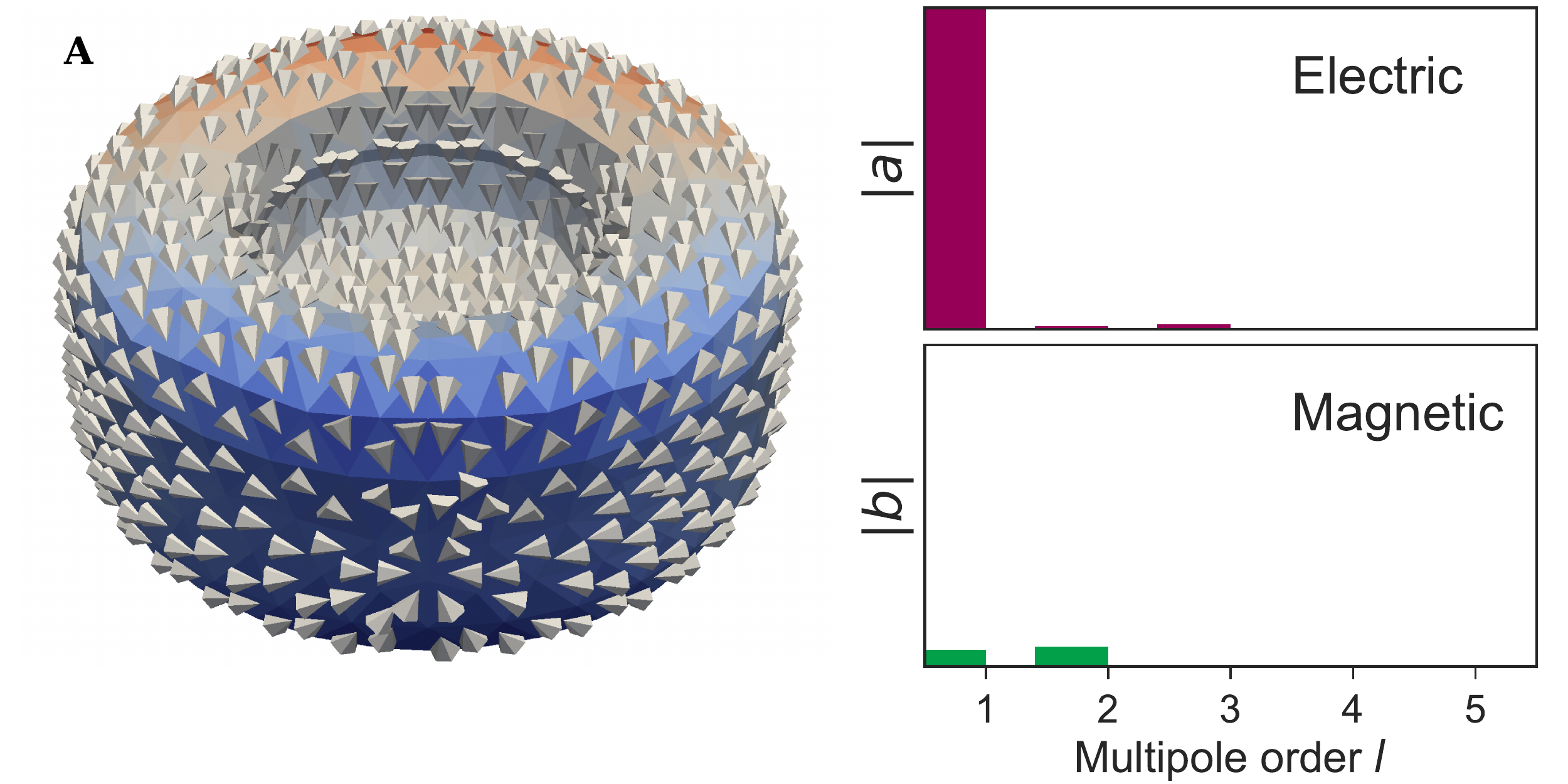}
\\[-10pt]\noindent\rule{0.94\columnwidth}{0.4pt}
\includegraphics[width=0.94\columnwidth]{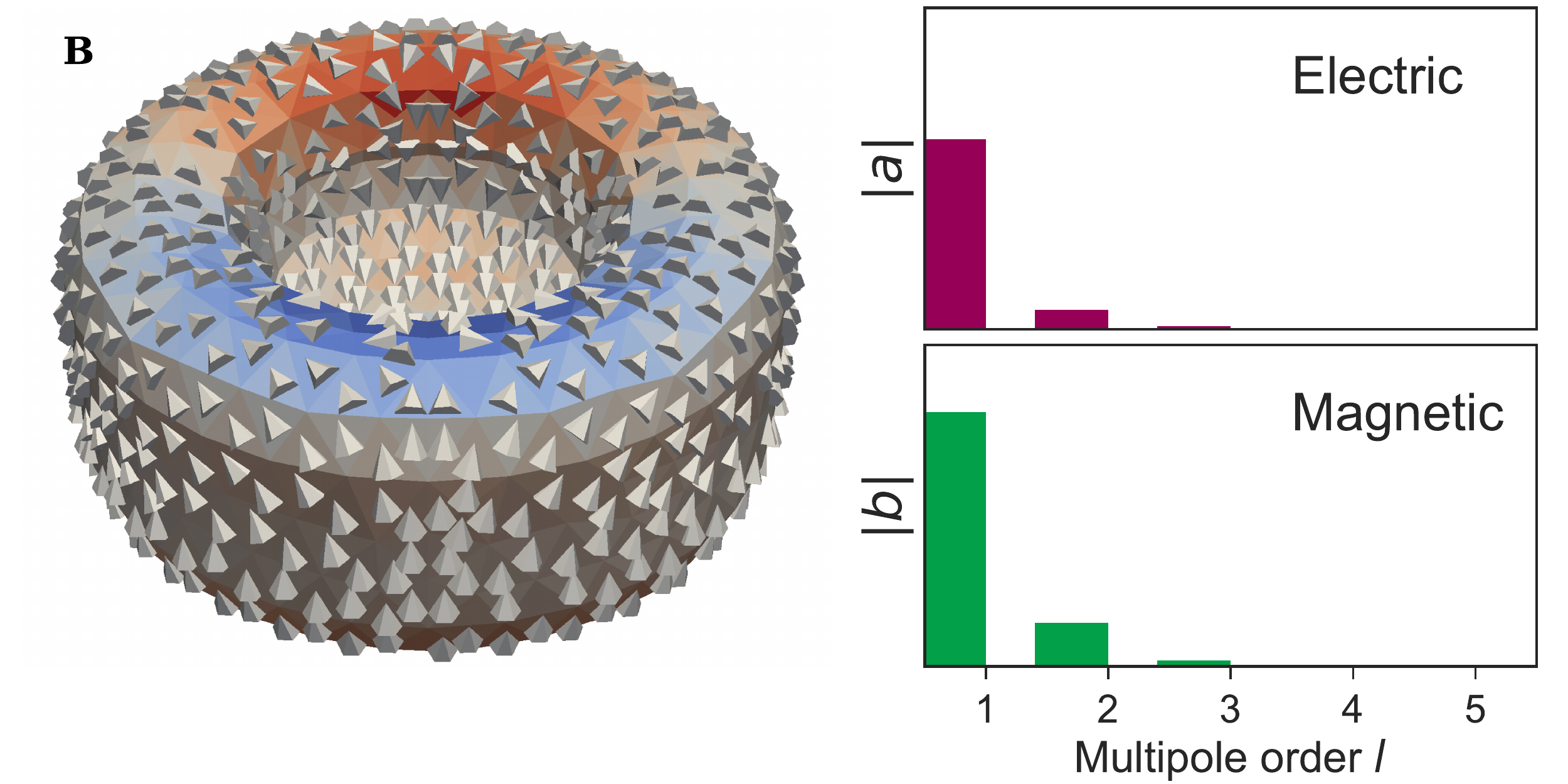}
\\[-10pt]\noindent\rule{0.94\columnwidth}{0.4pt}
\includegraphics[width=0.94\columnwidth]{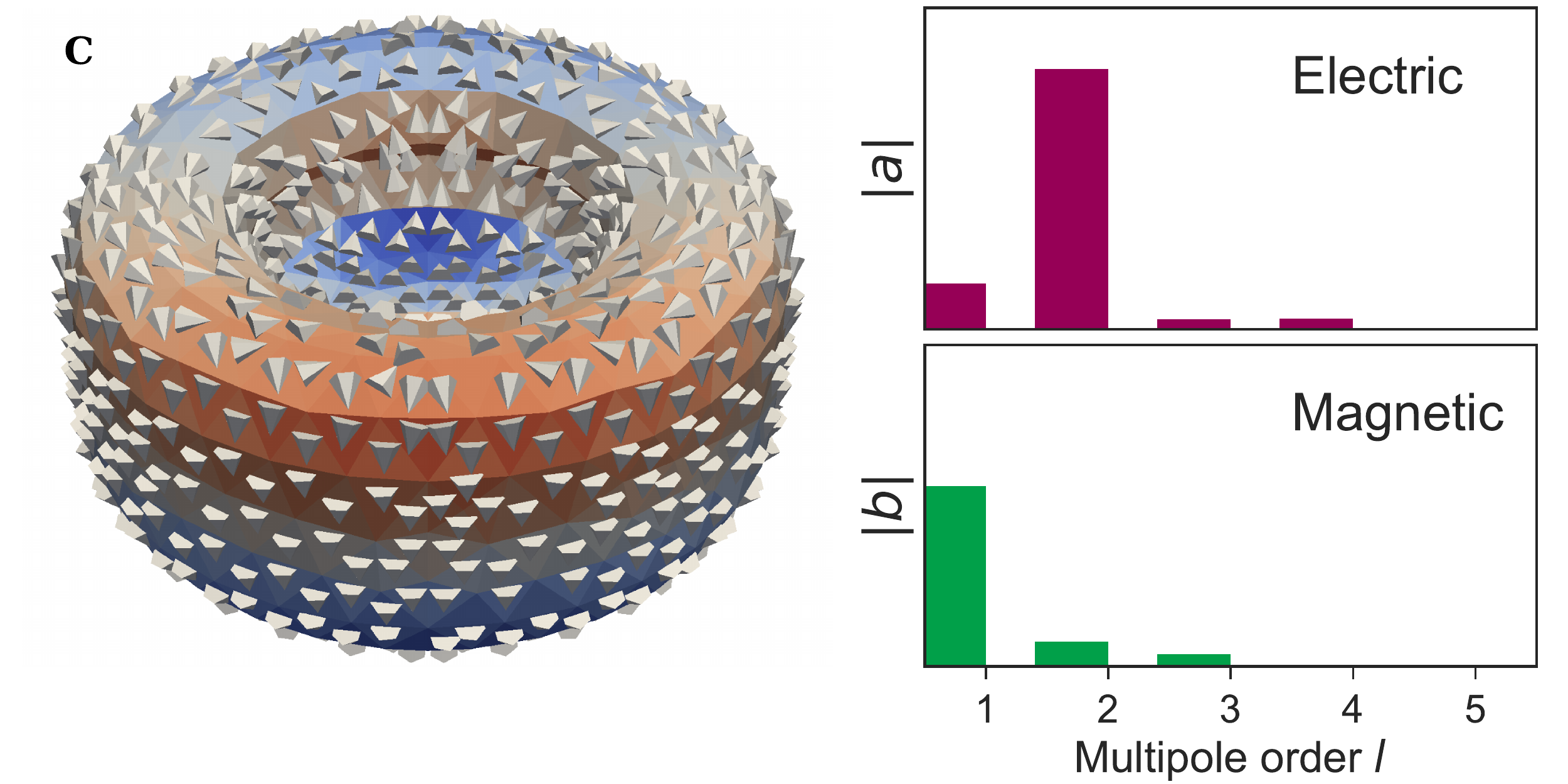}
\caption{Left: Modes of bianisotropic the disk with hole, showing equivalent electric surface currents (arrows), and charges (colours). Right: Spherical multipoles of each mode, normalised to the total scattered power.\label{fig:hole-mode-currents}}
\end{figure}

Recent theoretical \cite{alaee_all-dielectric_2015} and experimental\cite{odit_experimental_2016} work has shown that placing a hole asymmetrically in a dielectric disk creates an all-dielectric bianisotropic meta-atom. The magneto-electric polarisability of this structure leads to asymmetric back-scattering, however it is unclear how the various modes of the structure contribute to this process. The structure considered has the same dimensions as the disk studied in Section \ref{sec:disk}, with the addition of a hole having radius 121\,nm and depth 110\,nm. In Fig.~\ref{fig:hole-mode-currents}, the first 3 modes of this structure are plotted, along with their multipole expansions. Since the structure is strongly perturbed by the introduction of the hole, these modes can be understood as mixtures of several modes of the regular disk shown in Fig.~\ref{fig:mode-currents}.

Mode A is a predominantly electric-dipole type mode, and it can be seen that it has a very similar current distribution to mode I of the simple disk. Mode B has quite significant electric and magnetic dipole contributions. Examining the current distribution, it can be seen to have circulating current between the front and back faces, similar to mode II of the disk. However, the current in the hole has opposite direction to that on the rim, leading to a poloidal current distribution which strongly resembles mode III. Mode C is most closely related to mode IV of the simple disk, having a quadrupolar surface charge, but also having significant magnetic dipole moments.

For all modes, it can be seen that the introduction of the hole has increased the influence of higher-order multipoles, although both modes A and B remain dominated by dipole moments. For the chosen geometric parameters, mode B is the most important contributor to the bianisotropic response. Considering the case where modes A and B are dominant, mode A will suppresses the bianisotropy by adding electric dipole polarisability which is cross-coupled to a weak magnetic dipole excitation. Thus tuning the spectral overlap between these modes enables the net bianisotropy of the structure to be controlled.

\subsection{Elliptical cylinder}

\begin{figure}
\includegraphics[width=\columnwidth]{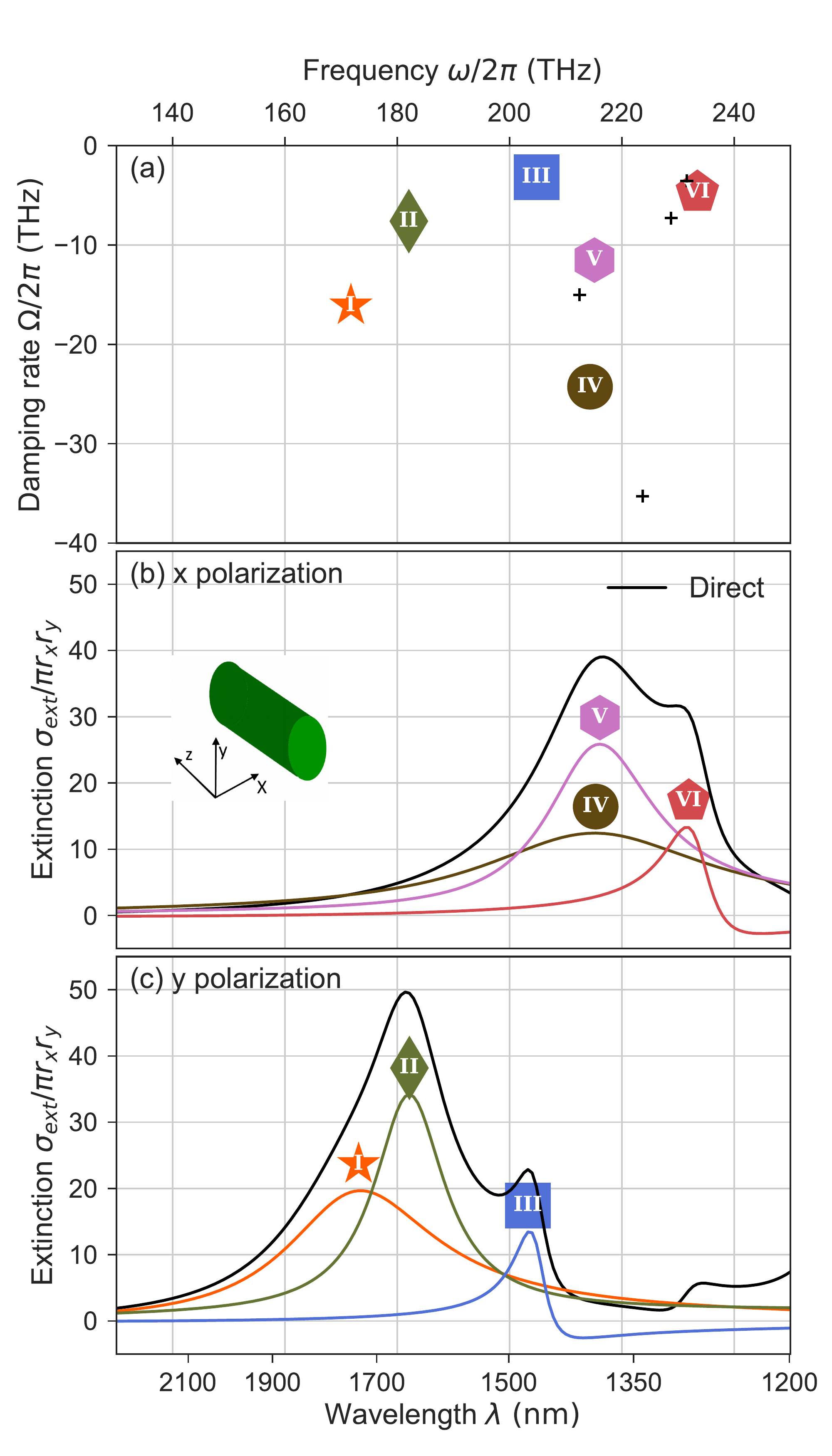}
\caption{\label{fig:elliptical_cylinder}(a) Location of poles for an elliptical silicon cylinder. (b) Extinction cross-section for the incident field polarised along the x axis (black solid line), along with contributions from the 3 dominant modes. The inset shows the coordinate convention. (c) Corresponding extinction for incident field polarized along the y axis.}
\end{figure}

Due to their rotational symmetry, the response of disks at normal incidence is identical for both polarizations. For applications in polarization manipulation, anisotropic structures are required. It has been demonstrated \cite{arbabi_dielectric_2015,kruk_invited_2016} that long dielectric cylinders of elliptical cross-section allow broadband birefringent metasurfaces to be fabricated, with applications as phase-plates, holograms and vector beam generators. It can be useful to think of such long meta-atoms as truncated sections of waveguide, where the transverse variation corresponds to a propagating waveguide mode.

\begin{figure*}[tb!]
\centering
\includegraphics[width=\columnwidth]{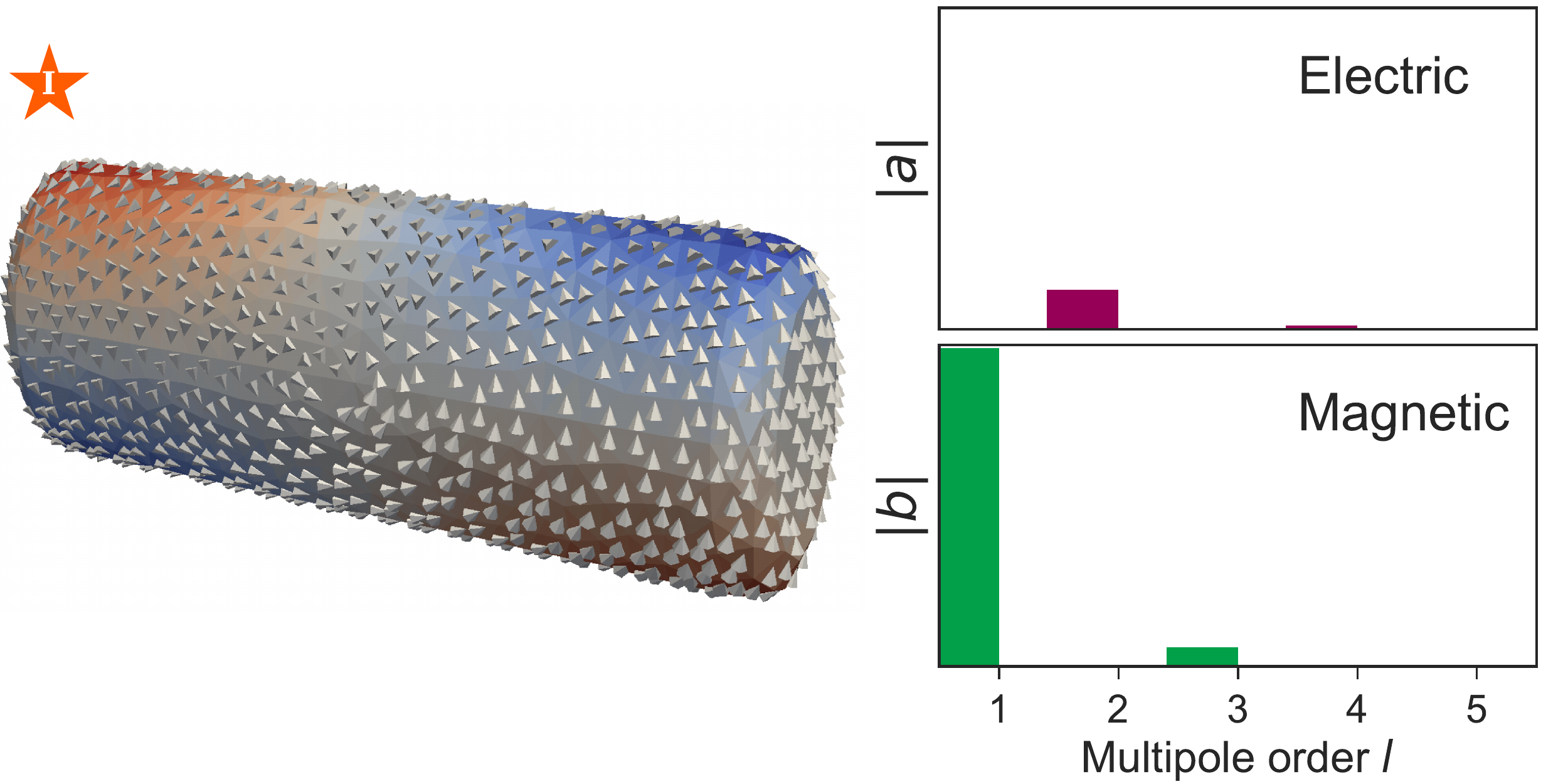}\includegraphics[width=\columnwidth]{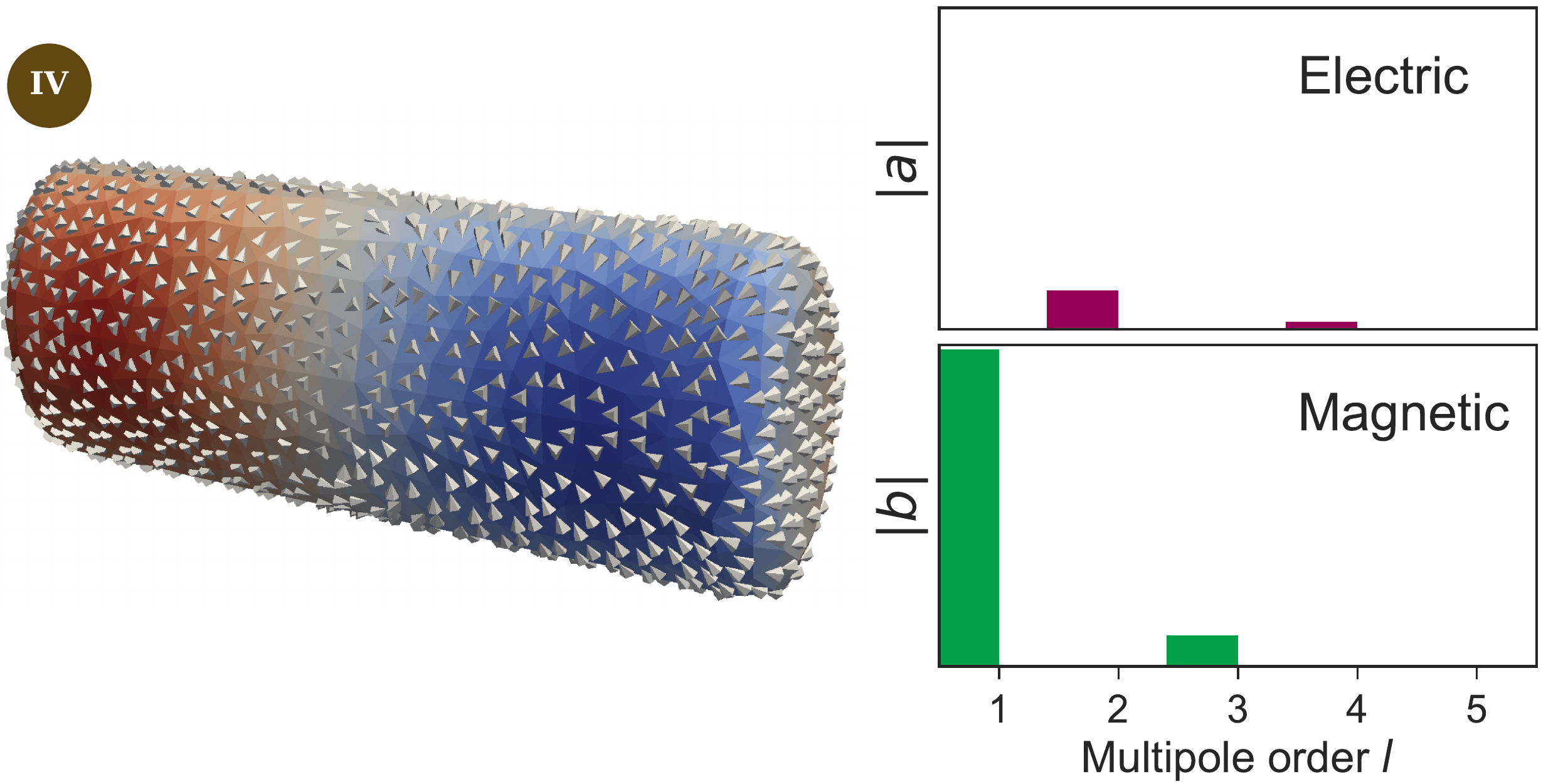}
\\[-10pt]\noindent\rule{\textwidth}{0.4pt}
\includegraphics[width=\columnwidth]{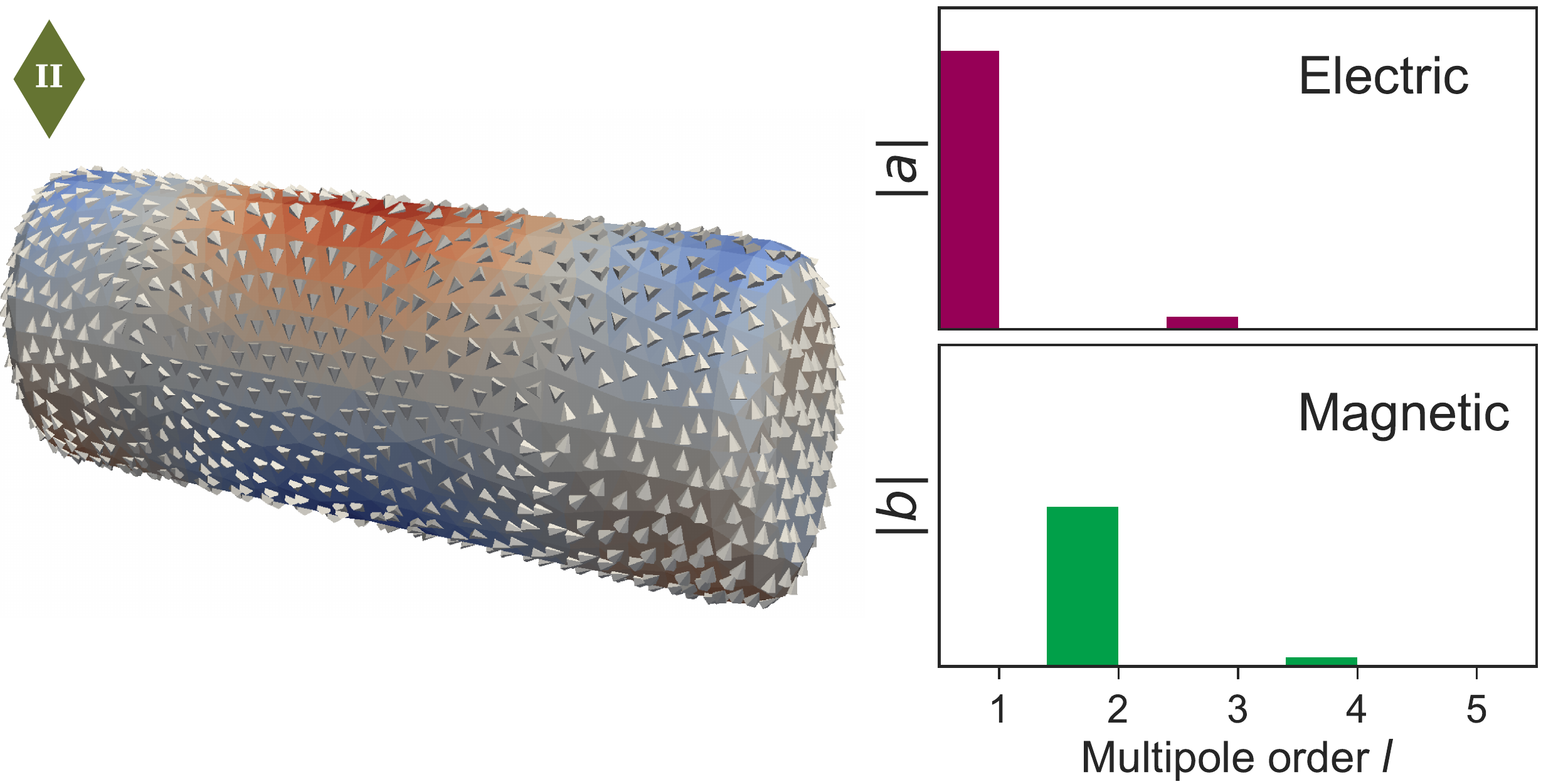}\includegraphics[width=\columnwidth]{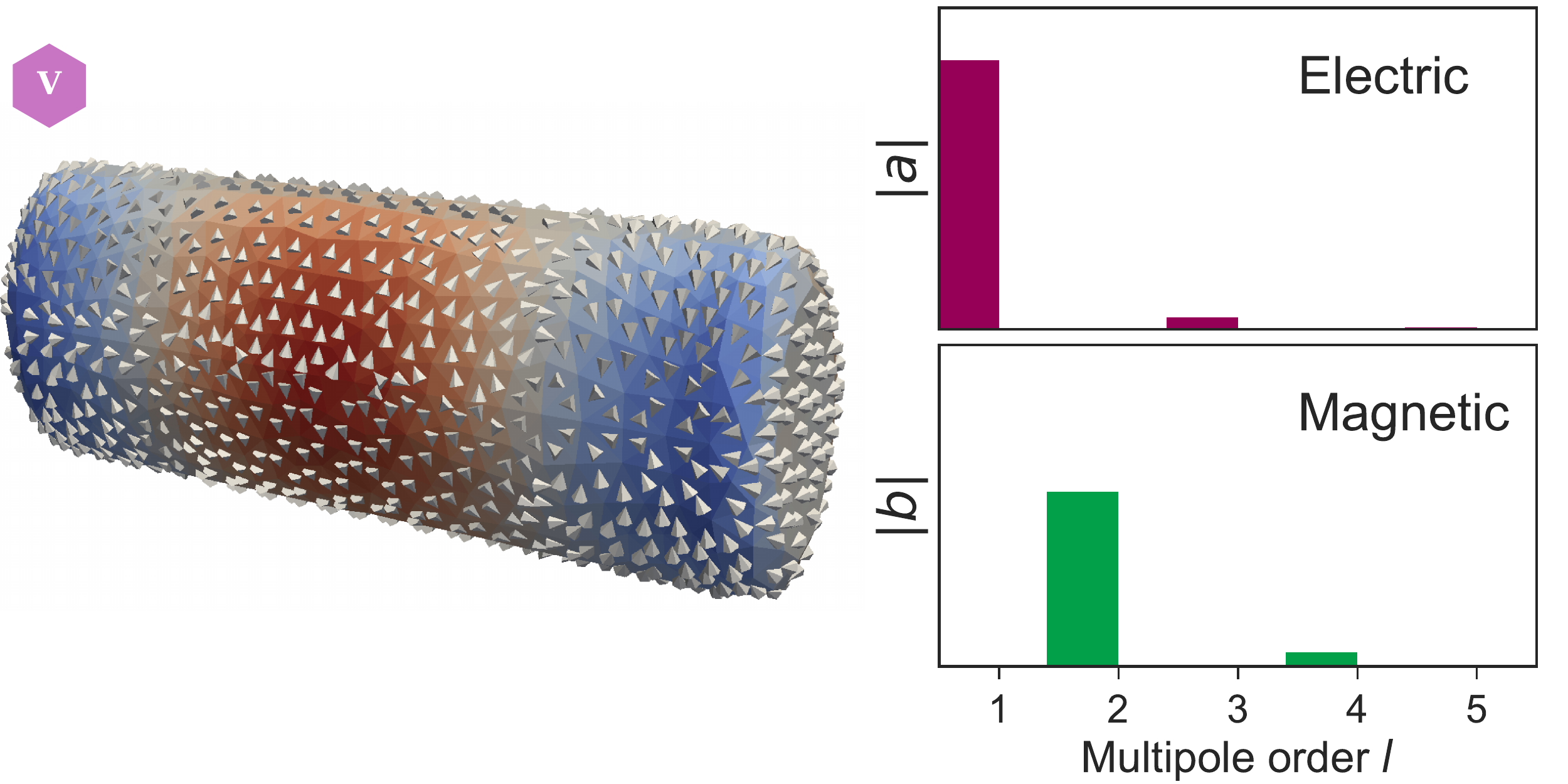}
\\[-10pt]\noindent\rule{\textwidth}{0.4pt}
\includegraphics[width=\columnwidth]{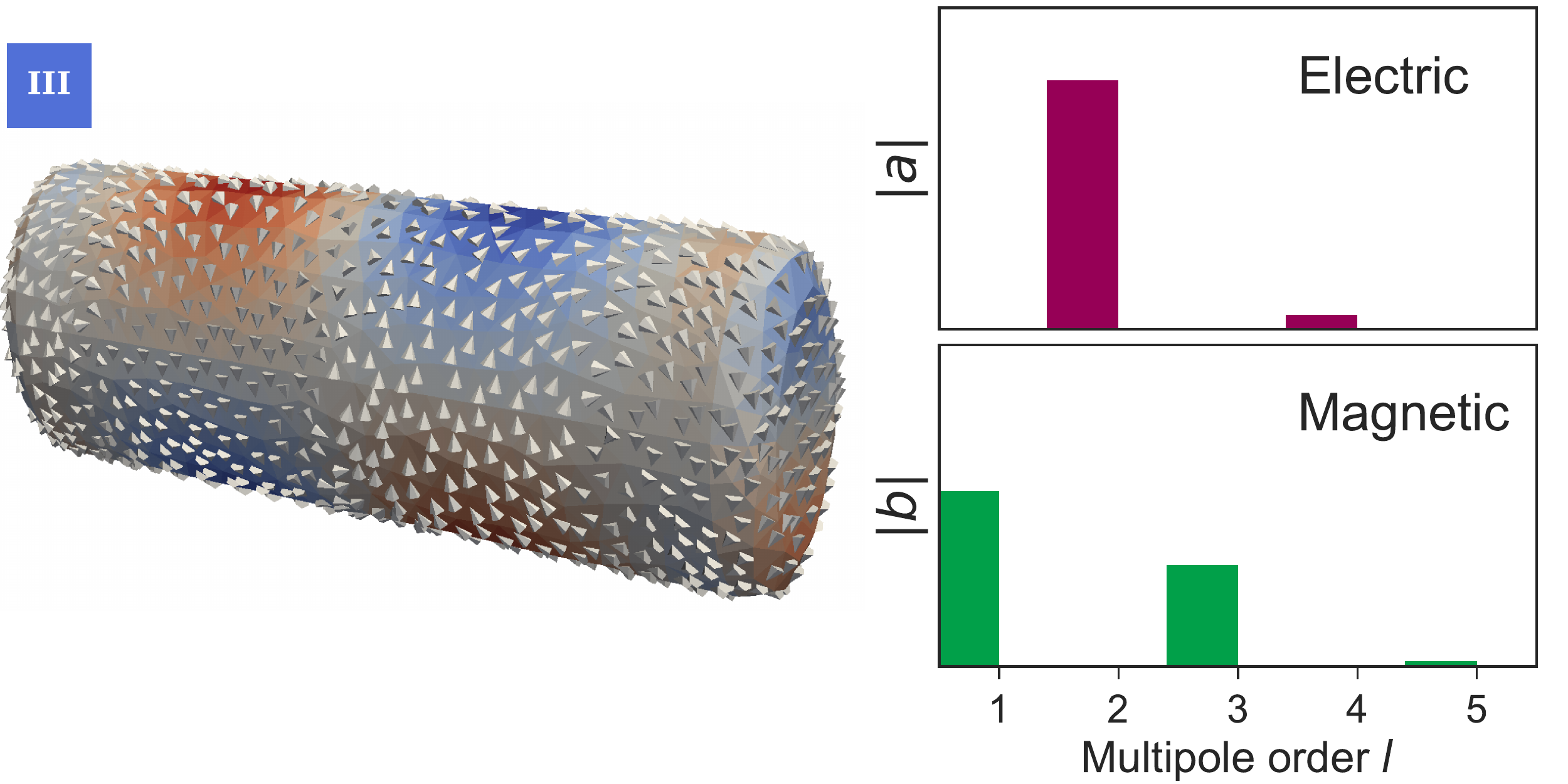}\includegraphics[width=\columnwidth]{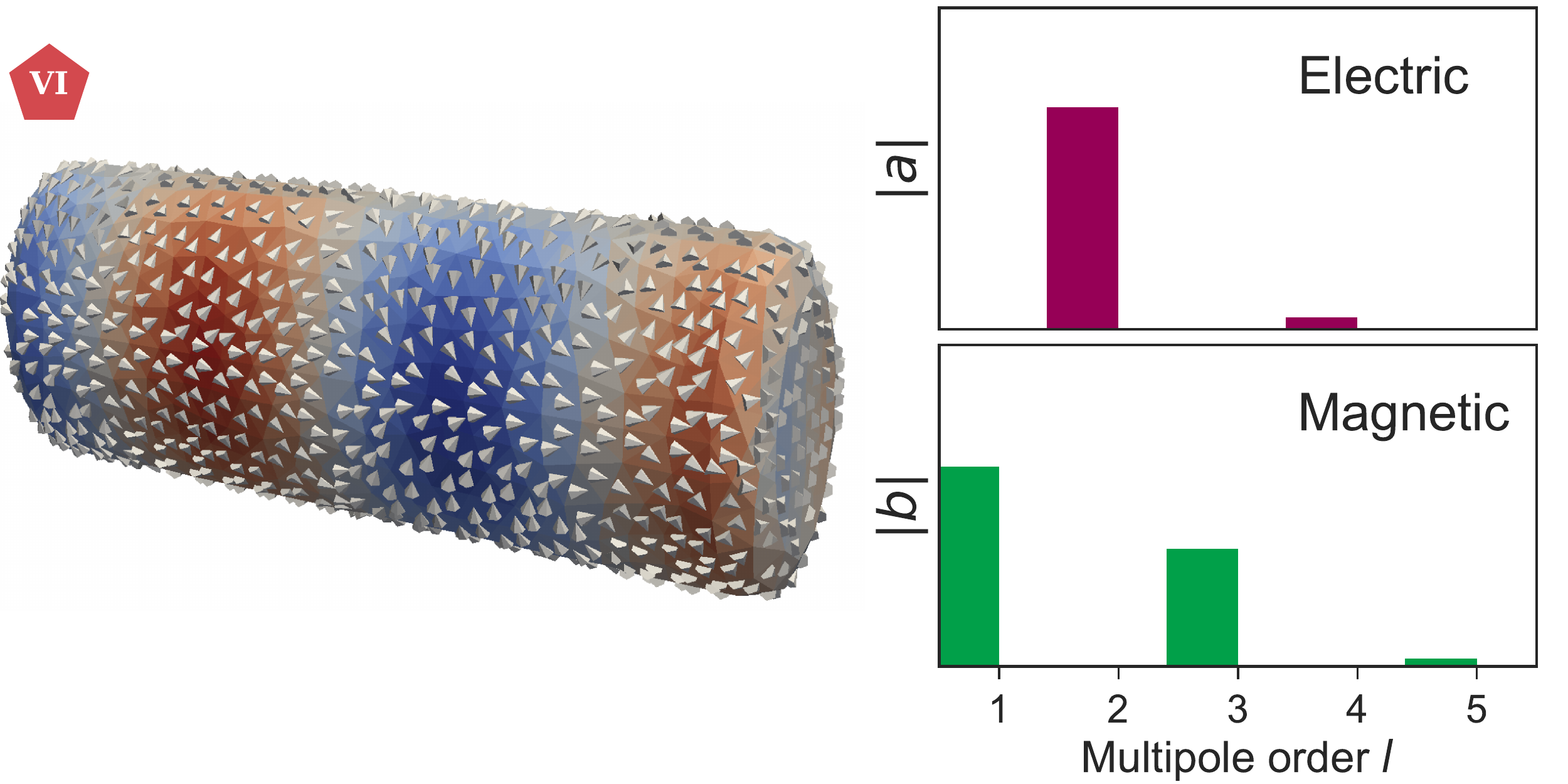}
\caption{Left: Modes of the elliptical cylinder, showing equivalent electric surface currents (arrows), and charges (colours). Markers correspond to poles in Fig.~\ref{fig:elliptical_cylinder}. Right: Spherical multipoles of each mode, normalised to the total scattered power.\label{fig:elliptical-mode-currents}}
\end{figure*}

In Fig.~\ref{fig:elliptical_cylinder}(a), the poles of an elliptical cylinder are shown, with x and y radii $r_x = 125$\,nm and $r_y = 200$\,nm and length $l=1100$\,nm. These parameters were chosen to approximately overlap several modes for both polarizations, to make forward scattering dominant. The corresponding surface current distributions and multipole moments are shown in Fig.~\ref{fig:elliptical-mode-currents}. In Fig.~\ref{fig:elliptical_cylinder}(b) and (c) the extinction is shown for $x$ and $y$ polarized incident plane waves respectively. The inset shows the coordinate convention. As expected the modes naturally divide into $x$ and $y$ polarizations, determined by the direction of surface currents on the incident face in Fig.~\ref{fig:elliptical-mode-currents}.

Examination of Fig.~\ref{fig:elliptical-mode-currents} shows that modes I and IV are both magnetic dipole-type, with currents circulating in the plane tangential to $\mathbf{H}$, accompanied by a quadrupolar surface charge distribution. Modes II and V are electric dipole type, with quite significant magnetic quadrupole contribution. Finally modes III and VI and dominated by their electric quadrupole moments, but also have quite significant magnetic dipole and octupole contributions. For each polarization, it can be seen that the higher-order modes have more field maxima in the longitudinal direction, but have comparable transverse field variations. This supports the idea that they arise from a few fundamental transverse waveguide modes, with different longitudinal variation corresponding to Fabry-Perot resonances. From Fig.~\ref{fig:elliptical_cylinder} it is clear that interference effects are considerably less pronounced in this elliptical cylinder than in the disk. This makes the structure simpler to analyze, but reduces the potential to tailor its spectral response by controlling interference.

\section{Conclusion}

A robust technique based on the singularity expansion method was presented to find the modes of a meta-atom, fully accounting for radiative losses. By solving Maxwell's equations using integral techniques, the problem of normalizing diverging fields is avoided. The technique was applied a silicon disk, a bianisotropic disk with hole, and an elliptical cylinder, which are all building blocks of experimentally demonstrated metasurfaces. It was demonstrated that the complicated features of the extinction spectrum can be readily explained in terms of contributions from the modes. Interference between non-orthogonal modes was shown to play a key role, and it was shown how the model automatically accounts for both direct and interference contributions to extinction.

When considering far-field scattering properties, a vector spherical harmonic expansion yields an accurate, if somewhat opaque, description. By combining it with the modal analysis, the nature and origin of all scattering features can be elucidated. In the case of the silicon disk, there are several bands of strong forward scattering and suppressed backscattering, corresponding to the generalised Huygens' condition. It was shown that each band corresponds to the overlap of modes with odd and even radiation symmetry. The techniques used to find modes and construct models of scatterers are implemented in an open-source code OpenModes\cite{powell_openmodes:_2014}, along with notebooks to reproduce all results in this paper\cite{powell_source_2016}.

\begin{acknowledgments}
The author acknowledges useful discussions with Andrey Miroshnichenko, Sarah Kostinski, Mingkai Liu, and Yuri Kivshar. This research was funded by the Australian Research Council.
\end{acknowledgments}

\appendix

\section{Integral approach to Maxwell's equations}\label{sec:integral_method}

Here a brief outline of the integral approach to solving Maxwell's equations is given. In this work dielectric objects are considered, and treated through a surface equivalent problem, with surface equivalent electric and magnetic currents, $\mathbf{J} = \mathbf{n}\times \mathbf{H}$ and $\mathbf{M} = -\mathbf{n}\times \mathbf{E}$, where $\mathbf{n}$ is the surface normal. These surface currents can be excited by the incident electric or magnetic field, yielding the electric field integral equation and magnetic field integral equation respectively. To yield a stable solution, both of these equations must be combined using some chosen weighting coefficients\cite{harrington_boundary_1989}. In this work the PMCHWT form is used\cite{yla-oijala_surface_2005}, which has been established to be positive-definite\cite{reid_efficient_2015,rodriguez_fluctuating-surface-current_2013}, as is required for a passive structure. This gives us an operator equation relating equivalent surface currents to the tangential components of the incident fields
\begin{equation} \label{eq:operator}
\mathcal{Z}\left(\mathbf{J}, \mathbf{M}\right) = \left.\left(\mathbf{E}_\mathrm{inc}, \mathbf{H}_\mathrm{inc}\right)\right|_\mathrm{tan}
\end{equation}

Equation \eqref{eq:operator} is solved numerically using the boundary element method (also known as the method of moments\cite{gibson_method_2008}). The equivalent surface currents are expanded in terms of a set of basis functions $\mathbf{f}_k(\mathbf{r})$
\begin{equation} \label{eq:current_expansion}
\mathbf{J}(\mathbf{r}) = \sum_{k=1}^{N}I_{k}\mathbf{f}_k(\mathbf{r}), \qquad
\mathbf{M}(\mathbf{r}) = \frac{1}{\eta_0}\sum_{k=1}^{N}I_{k+N}\mathbf{f}_k(\mathbf{r}),
\end{equation}
where $\eta_0$ is the impedance of free space. The current weighting coefficients $I_k$ are assembled into the vector $\mathbf{I}$.

The current expanded in terms of a finite series of basis functions as per Eq.~\eqref{eq:current_expansion} cannot exactly satisfy Eq.~\eqref{eq:operator}. Therefore, it must be solved by minimizing the residual error with respect to some weighting functions $\mathbf{g}_k$. These are applied to the source fields, yielding source coefficients
\begin{equation}
V_k = \left[\int\mathbf{g}_k(\mathbf{r})\cdot\mathbf{E}_\mathrm{inc}(\mathbf{r})d^2\mathbf{r}, \eta_0\int\mathbf{g}_k(\mathbf{r})\cdot\mathbf{H}_\mathrm{inc}(\mathbf{r})d^2\mathbf{r}\right]^T,
\end{equation}
which are assembled into the source vector $\mathbf{V}$. In this work loop-star functions \cite{vecchi_loop-star_1999} are used for both basis and testing functions, since using the more common RWG\cite{rao_electromagnetic_1982} first order linear functions was found to generate many spurious poles. The weighted operator $\mathcal{Z}$ has a complex expression which can be found in Ref.~\onlinecite{yla-oijala_surface_2005}, resulting in the impedance matrix $\mathbf{Z}(s)$.

The response of the system is now described by a matrix equation,
\begin{equation}
\mathbf{V}(s)=\mathbf{Z}(s)\cdot\mathbf{I}(s),
\end{equation}
The impedance matrix $\mathbf{Z}(s)$ is dense and frequency dependent, and contains all information regarding the response of the scatterer to arbitrary incident fields.

\section{Poles of the impedance matrix} \label{sec:pole_search}

It can be seen from Eq.~\eqref{eq:impedance} that the singularities of $\mathbf{Z}^{-1}(s)$ will dominate the spectrum of the response, and by Mittag-Leffler's theorem the response may be expanded in terms of these singularities \cite{pearson_evidence_1981}. They correspond to solutions which can exist in the absence of a source, and hence they can be used to model the response to an arbitrary incident field. The most important singularities of the impedance matrix are its poles, corresponding to the quasi-normal modes of the system. In practice it may usually be assumed that all poles are of first order \cite{marin_major_1981}. 

The poles of the impedance matrix are found by the contour integration procedure of Ref.~\onlinecite{bykov_numerical_2013}. First a pair of matrix integrals $\mathbf{C}_1 = \oint \mathbf{Z}^{-1}(s)ds$ and $\mathbf{C}_2 = \oint s \mathbf{Z}^{-1}(s)ds$ is evaluated about a contour containing all modes of interest, as shown schematically in Fig.~\ref{fig:contour}. As discussed in Section \ref{sec:defining-modes}, the contour is chosen to enclose only those modes which are likely to be of physical interest. Also note that an arc is used to eliminate spurious numerical poles which cluster near the origin when using integral operators of the first kind \cite{hanson_operator_2002}.

The mode frequencies and currents are eigenvalues and eigenvectors of $\mathbf{C}_2 \cdot \mathbf{I}_n = s_n \mathbf{C}_1 \cdot \mathbf{I}_n$. A singular value decomposition is used to determine the number of valid solutions to this equation \cite{bykov_numerical_2013} and solving for the corresponding left eigenvalue problem yields the projectors $\mathbf{K}_n$. This procedure can yield solutions lying both inside and outside the contour, and those falling outside the contour are discarded. The poles and currents are further improved by Newton iteration, then normalised so that $\mathbf{K}_n \cdot \mathbf{Z}'(s_n) \cdot \mathbf{I}_n=1$. 
This ensures that the dyadic product of the eigenvectors matches the pole residue, i.e.
\begin{equation} \label{eq:residue}
\mathbf{Z}(s)=\frac{\mathbf{I}_n\mathbf{K}_n}{s-s_n},
\end{equation}
%
in the vicinity of $s_n$, simplifying the pole expansion. 

When solving the structure numerically, the imperfect symmetry of the mesh usually results in some frequency splitting of degenerate modes, so a thresholding procedure is used to group closely spaced poles. The contour integration and iterative search procedure were found to cope with these nearly degenerate poles without requiring any special handling. Note that it is not necessary to orthogonalise degenerate modes, since the method is intrinsically able to account for non-orthogonality, as long as the modes span the full eigenspace.

\section{Orthogonality of the modes} \label{sec:orthogonality}

As discussed in Ref.~\onlinecite{leung_completeness_1994}, the electric fields of quasi-normal modes do not obey the usual orthogonality relationship based on a conjugated inner product, i.e. $\int \mathbf{E}_n^{*}\cdot \mathbf{E}_m d^3\mathbf{r} \neq \delta_{nm}$. However, they do obey an unconjugated orthgonality relationship, which is utilized in most quasi-normal mode formulations\cite{kristensen_normalization_2015} for normalization of modes, and for projection of external fields.

In contrast, the current vectors on the scatterer obtained in this work do not exhibit any form of orthogonality. Such orthogonality is not required when working with modal currents, since they are normalized by weighting them to match the residue of the pole, as shown in Eq.~\eqref{eq:residue}. In addition to providing the current vector $\mathbf{I}_n$, this approach also yields the correctly normalised projector $\mathbf{K}_n$, which gives the projection of an arbitrary field onto each mode by a simple scalar product, as used in Eq.~\eqref{eq:current_modes}.

It is noted that in the literature a number of orthogonal decompositions of the impedance matrix $\mathbf{Z}$ have been presented, most prominently the characteristic mode analysis \cite{lau_guest_2016}. As these mode vectors are real, they exhibit the conventional conjugated orthogonality. However, such decompositions suffer from a number of problems which make them unsuited for physically modelling open resonators. First, the eigenvalue problem must be solved at each frequency, yielding a different set of current vectors at each frequency. This requires some algorithm to track modes with frequency \cite{safin_advanced_2016}, and effectively prevents their use in time-domain problems.

More significantly, the enforcement of mode orthogonality on an inherently non-Hermitian system results in an artificial set of basis vectors which contain a complex mixture of underlying eigenvectors. This manifests itself in unphysical avoided crossings, whereby the nature of a pair of modes is swapped in some frequency region \cite{schab_eigenvalue_2016}. The author has observed similar behaviour when utilising other orthogonal decompositions of the impedance matrix, such as the singular value decomposition. In order to reproduce the interference phenomena observed in Fig.~\ref{fig:extinction_modes}, it is essential to use the non-orthgonal modes obtained from the singularity expansion method, or quasi-normal mode approaches.

\section{Multipole decomposition} \label{sec:multipole}

The electric multipole coefficients $a_{lm}$ and magnetic multipoles coefficients $b_{lm}$ were computed directly from the surface currents using the formulas from Ref.~\onlinecite{grahn_electromagnetic_2012}. Duality allows these formulas to be generalized to include the equivalent magnetic currents through the substitution $\mathbf{J} \rightarrow j\frac{1}{\eta_0}\mathbf{M}$. The normalization of multipole coefficients from Ref.~\onlinecite{jackson_classical_1999} is used, as this simplifies the expression for the scattering cross-section, which is given by
\begin{equation} \label{eq:multipole-scattering}
\sigma_{scat} = \frac{\eta_0}{k^2|E_0|^2}\sum_{l=1}^{l_{max}}|a_{l}|^2+|b_{l}|^2,
\end{equation}
where the coefficients include contributions from all values of azimuthal index $m$:
\begin{equation}
|a_l|^2 = \sum_{m=-l}^{l}|a_{lm}|^2, \qquad |b_l|^2 = \sum_{m=-l}^{l}|b_{lm}|^2.
\end{equation}


%
In Fig.~\ref{fig:mode-currents} $|a_l|^2$ and $|b_l|^2$ are normalised to their sum, and their square root is plotted since it more clearly shows the smaller contributions. In Fig.~\ref{fig:multipole-scattering} these terms are plotted including the pre-factor from Eq.~\eqref{eq:multipole-scattering} to give them dimensions of scattering cross-section.

For a plane wave propagating in the $z$ direction, with incident electric field along the $y$ direction, the extinction cross-section is given by \cite{grahn_electromagnetic_2012}
\begin{eqnarray}
\sigma_{ext} = \frac{\pi}{k^2}\sum_{l=1}^{l_{max}}\sqrt{2l+1}\left(\left[\sum_{m=-1,1}\mathrm{Im}\{a_{lm}\}\right]\right.\nonumber\\*
+ \left.\left[\sum_{m=-1,1}m\mathrm{Im}\{b_{lm}\}\right]\right). \label{eq:multipole_extinction}
\end{eqnarray}
The quantities in square brackets are plotted in Fig.~\ref{fig:extinction_multipole}, including all common pre-factors in Eq.~\eqref{eq:multipole_extinction}.
For 3 terms of the multipole expansion, the extinction plotted in Fig.~\ref{fig:extinction_multipole} agrees with the direct calculation to a relative error below 2\% for frequencies below 350THz.

By adapting the formulas from Mie theory \cite{bohren_absorption_1983}, forward scattering can be found as
\begin{equation}
W_{f} = \frac{\pi}{4k^2}\left|\sum_{l=1}^{l_{max}}\sqrt{2l+1}\sum_{m=-1,1}^{l}a_{lm}+mb_{lm}\right|^2.
\end{equation}
while back-scattering is given by
\begin{equation}
W_{b} = \frac{\pi}{4k^2}\left|\sum_{l=1}^{l_{max}}(-1)^l\sqrt{2l+1}\sum_{m=-1,1}^{l}a_{lm}-mb_{lm}\right|^2.
\end{equation}

As losses are low in this system, the total extinction and scattering are approximately equal, due to the optical theorem. However, this still allows each multipole's contribution to extinction shown in Fig.~\ref{fig:extinction_multipole} to be different from its contribution to scattering shown in Fig.~\ref{fig:multipole-scattering}.

\section{Accuracy of the modal expansion} \label{sec:accuracy}

\begin{figure}[htb]
\includegraphics[width=\columnwidth]{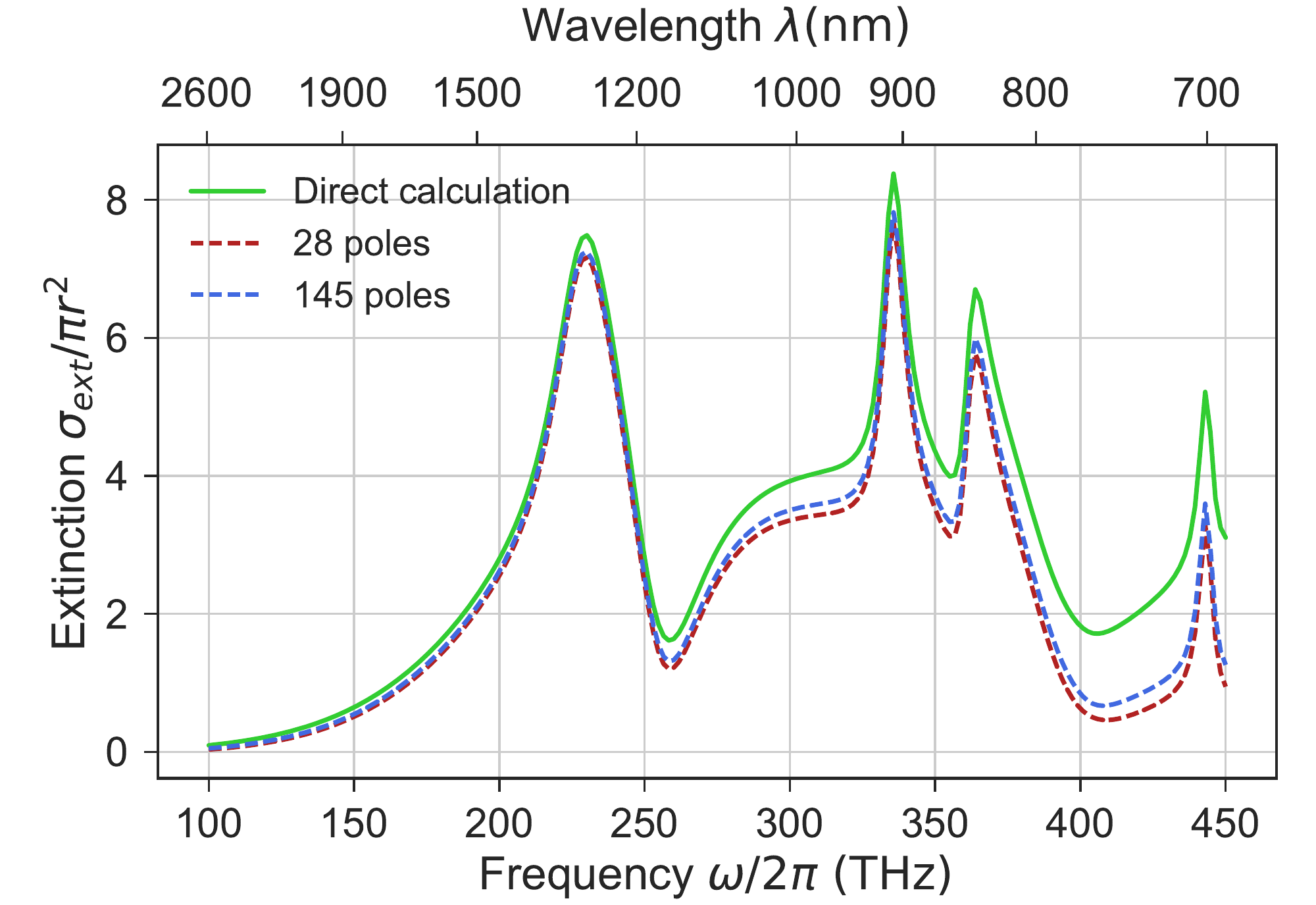}
\caption{Accuracy of the extinction calculated from the model including different numbers of modes (dashed lines), compared the with directly calculated result (solid line). \label{fig:extinction-accuracy}}
\end{figure}

To confirm the accuracy of the mode expansion, the directly calculated extinction curve is plotted in Fig.~\ref{fig:extinction-accuracy} (solid line), as well as the sum of all contributions plotted in Fig.~\ref{fig:extinction_modes}(b) (red dashed line). It can be seen that the agreement is good for frequencies below 250THz, however at high frequencies it becomes poorer. In this curve the number of poles considered is 28, corresponding to the 7 modes studied in Section \ref{sec:disk}, each doubly degenerate and with conjugate poles.

To improve agreement all 145 poles found by the contour integration process are included, not just the most significant. In this case, some included modes are polarized along the cylinder axis, and hence are not doubly degenerate, while others are over-damped, and hence do not appear in conjugate pairs. Including all of these poles, much better agreement is achieved, as shown by the blue dashed curve. Clearly a model involving so many parameters is less useful as a design tool, thus there is an inevitable trade-off between accuracy and the level of insight provided. However, in contrast to simpler approaches based on point dipole or equivalent circuit models, it is possible to control the level of detail which is included within the model by choosing to include or exclude poles.

\section{Material models} \label{sec:materials}

Any material may be incorporated into the model, as long as its permittivity (and permeability if applicable) can be described by a meromorphic function in the complex frequency plane. This corresponds to the permittivity having a real, causal representation in the time-domain, and is exactly the same issue faced when creating material models for use with e.g.~the finite-difference time-domain method. The common material model of a sum of Drude-Lorentz oscillators, or related approaches can be used \cite{han_model_2006}. Such material models simultaneously include both dispersion effects, as required by the Kramers-Kronig relations. Except in the case of idealized lossless and non-dispersive materials, the material permittivity will have its own poles and zeros in the complex plane. In general, the pole frequencies of a scatterer are influenced both by its geometry, and by the poles and zeros of the material permittivity.

It is important to note that the the impedance matrix $\mathbf{Z}(s)$ contains terms of the form $\exp(-\gamma_0r)$, with complex wave-number $\gamma_0 = \sqrt{\varepsilon(s)}\frac{s}{c}$. Due to the presence of the square root operation, this results in branch points at the poles and zeros of the permittivity, connected by branch cuts\cite{makitalo_modes_2014,mansuripur_leaky_2016}. For the material data used in this work, all such branch points occur at frequencies above 800\,THz, thus their contribution is neglected in Eq.~\eqref{eq:current_modes}. The accuracy of the results shown in Fig.~\ref{fig:extinction-accuracy} confirms that no significant contribution from branch points is missing from the result. The lack of branch points in the frequency range of interest also ensures that the integration contour illustrated in Fig.~\ref{fig:contour} does not intersect any of the branch cuts.  Applying the contour integration in a frequency range of high material dispersion would require choosing the contour carefully to account for all branch-cuts.

The surface equivalence approach used in this work is most suited to structures composed of a single material. It can be extended to multi-material structures by also solving for equivalent surface currents on the internal boundary between materials \cite{yla-oijala_application_2005}. For composite particles with highly-complex internal structure, the volume integral approach of Ref.~\onlinecite{zheng_implementation_2014} may be preferable.

%

\end{document}